\documentclass[12pt]{article}

\begin{document}

\def\CA{{\cal A}}
\def\CB{{\cal B}}
\def\CC{{\cal C}}
\def\CD{{\cal D}}
\def\CE{{\cal E}}
\def\CF{{\cal F}}
\def\CG{{\cal G}}
\def\CH{{\cal H}}
\def\CI{{\cal I}}
\def\CJ{{\cal J}}
\def\CK{{\cal K}}
\def\CL{{\cal L}}
\def\CM{{\cal M}}
\def\CN{{\cal N}}
\def\CO{{\cal O}}
\def\CP{{\cal P}}
\def\CQ{{\cal Q}}
\def\CR{{\cal R}}
\def\CS{{\cal S}}
\def\CT{{\cal T}}
\def\CU{{\cal U}}
\def\CV{{\cal V}}
\def\CW{{\cal W}}
\def\CX{{\cal X}}
\def\CY{{\cal Y}}
\def\CZ{{\cal Z}}

\newcommand{\todo}[1]{{\em \small {#1}}\marginpar{$\Longleftarrow$}}
\newcommand{\labell}[1]{\label{#1}\qquad_{#1}} 
\newcommand{\bbibitem}[1]{\bibitem{#1}\marginpar{#1}}
\newcommand{\llabel}[1]{\label{#1}\marginpar{#1}}

\newcommand{\sphere}[0]{{\rm S}^3}
\newcommand{\su}[0]{{\rm SU(2)}}
\newcommand{\so}[0]{{\rm SO(4)}}
\newcommand{\bK}[0]{{\bf K}}
\newcommand{\bL}[0]{{\bf L}}
\newcommand{\bR}[0]{{\bf R}}
\newcommand{\tK}[0]{\tilde{K}}
\newcommand{\tL}[0]{\bar{L}}
\newcommand{\tR}[0]{\tilde{R}}

\newcommand{\btzm}[0]{BTZ$_{\rm M}$}
\newcommand{\ads}[1]{{\rm AdS}_{#1}}
\newcommand{\ds}[1]{{\rm dS}_{#1}}
\newcommand{\eds}[1]{{\rm EdS}_{#1}}
\newcommand{\sph}[1]{{\rm S}^{#1}}
\newcommand{\gn}[0]{G_N}
\newcommand{\SL}[0]{{\rm SL}(2,R)}
\newcommand{\cosm}[0]{R}
\newcommand{\hdim}[0]{\bar{h}}
\newcommand{\bw}[0]{\bar{w}}
\newcommand{\bz}[0]{\bar{z}}
\newcommand{\be}{\begin{equation}}
\newcommand{\ee}{\end{equation}}
\newcommand{\bea}{\begin{eqnarray}}
\newcommand{\eea}{\end{eqnarray}}
\newcommand{\pat}{\partial}
\newcommand{\lp}{\lambda_+}
\newcommand{\bx}{ {\bf x}}
\newcommand{\bk}{{\bf k}}
\newcommand{\bb}{{\bf b}}
\newcommand{\BB}{{\bf B}}
\newcommand{\tp}{\tilde{\phi}}
\hyphenation{Min-kow-ski}

\def\apr{\alpha'}
\def\str{{str}}
\def\lstr{\ell_\str}
\def\gstr{g_\str}
\def\Mstr{M_\str}
\def\lpl{\ell_{pl}}
\def\Mpl{M_{pl}}
\def\varep{\varepsilon}
\def\del{\nabla}
\def\grad{\nabla}
\def\tr{\hbox{tr}}
\def\perp{\bot}
\def\half{\frac{1}{2}}
\def\p{\partial}
\def\perp{\bot}
\def\eps{\epsilon}

\def\NPB{{\it Nucl. Phys. }{\bf B}}
\def\PL{{\it Phys. Lett. }}
\def\PRL{{\it Phys. Rev. Lett. }}
\def\PRD{{\it Phys. Rev. }{\bf D}}
\def\CQG{{\it Class. Quantum Grav. }}
\def\JMP{{\it J. Math. Phys. }}
\def\SJNP{{\it Sov. J. Nucl. Phys. }}
\def\SPJ{{\it Sov. Phys. J. }}
\def\JETPL{{\it JETP Lett. }}
\def\TMP{{\it Theor. Math. Phys. }}
\def\IJMPA{{\it Int. J. Mod. Phys. }{\bf A}}
\def\MPL{{\it Mod. Phys. Lett. }}
\def\CMP{{\it Commun. Math. Phys. }}
\def\AP{{\it Ann. Phys. }}
\def\PR{{\it Phys. Rep. }}

\renewcommand{\thepage}{\arabic{page}}
\setcounter{page}{1}

\rightline{hep-th/0207245}
\rightline{VPI-IPPAP-02-05, UPR-1008-T,
ITFA-2002-26}
\vskip 0.75 cm
\centerline{\Large \bf Exploring de Sitter Space and Holography}
\vskip 0.75 cm

\renewcommand{\thefootnote}{\fnsymbol{footnote}}
\centerline{{\bf Vijay
Balasubramanian,${}^{1}$\footnote{vijay@endive.hep.upenn.edu}
Jan de Boer,${}^{2}$\footnote{jdeboer@wins.uva.nl}
and
Djordje Minic${}^{3}$\footnote{dminic@vt.edu}
}}
\vskip .5cm
\centerline{${}^1$\it David Rittenhouse Laboratories, University of
Pennsylvania}
\centerline{\it Philadelphia, PA 19104, U.S.A.}
\vskip .5cm
\centerline{${}^2$\it Instituut voor Theoretische Fysica,}
\centerline{\it Valckenierstraat 65, 1018XE Amsterdam, The Netherlands}
\vskip .5cm
\centerline{${}^3$\it Institute for Particle Physics and Astrophysics}
\centerline{\it Department of Physics, Virginia Tech}
\centerline{\it Blacksburg, VA 24061, U.S.A.}
\vskip .5cm

\setcounter{footnote}{0}
\renewcommand{\thefootnote}{\arabic{footnote}}

\begin{abstract}
We explore aspects of the physics of de Sitter (dS)  space that are relevant to holography with a positive cosmological constant.  First we display a nonlocal map that commutes with the de Sitter isometries, transforms the bulk-boundary propagator and solutions of free wave equations in de Sitter onto the same quantities in Euclidean anti-de Sitter (EAdS), and takes the two boundaries of dS to the single EAdS boundary via an antipodal identification.  Second we compute the action of scalar fields on dS as a functional of boundary data.    Third, we display a family of solutions to 3d gravity with a positive cosmological constant in which the equal time sections are arbitrary genus Riemann surfaces, and compute the action of these spaces as a functional of boundary data from the Einstein gravity and Chern-Simons gravity points of view.   These studies suggest that if de Sitter space is dual to a Euclidean conformal field theory (CFT), this theory should involve two disjoint, but possibly entangled factors.   We argue that these CFTs would be of a novel form, with unusual hermiticity conditions relating left movers and right movers.   After exploring these conditions in a toy model, we combine our observations to propose that a holographic dual description of de Sitter space would involve a pure entangled state in a product of two of our unconventional CFTs associated with the de Sitter boundaries.   This state can be constructed to preserve the de Sitter symmetries and and its decomposition in a basis appropriate to antipodal inertial observers would lead to the thermal properties of static patch.   
\end{abstract}


\section{Introduction}

Recent work has studied the possibility that de Sitter (dS) space
is holographically dual to a Euclidean field theory defined on the
late and early time conformal boundaries ($\CI^\pm$)  of this
universe (see, e.g.,  \cite{hull, bhmds, wittds, andyds, klemm1, bdbmds,
strombousso}).    While the status of the duality remains unclear,
exploration of this possiblity has led to many interesting new
results.  Amongst other successes,  we have learned much about the
asymptotic symmetries of de Sitter space (e.g., see \cite{andyds,
klemm1, bdbmds}), we now have a method of measuring the
gravitational mass of asymptotically de Sitter universes (e.g.,
\cite{klemm1, bdbmds,mann,moremass}), and we have understood new aspects of the
famous ambiguity in choosing in the vacuum state of fields in de
Sitter space~\cite{strombousso}.  These developments have been
aided by a simple fact:  de Sitter space ($\Lambda > 0$) and
Euclidean AdS space (EAdS, $\Lambda < 0$) are both hyperboloids
\begin{equation}
-x_0^1 + x_1^2 + \cdots + x_d^2 = l^2 ~~~;~~~ |\Lambda| = {(d-1)(d-2) \over (2l^2)}
\label{hyperboloid}
\end{equation}
embedded in flat space with signature (1, d).  As a result,
analytic continuation of various facts about classical AdS leads
to facts about classical dS -- for example, we can learn that the
asymptotic symmetry group of dS like AdS is the conformal group
following~\cite{andyds,bh,fefgrah,hensken,brownyork,stressvp,mott,nojiri}.  However, the quantum mechanical physics of de Sitter space does not follow from
analytic continuation -- for example, the Green function for a
scalar field obtained by analytic continuation from Euclidean AdS
is the not the two point function of the scalar field in a de
Sitter vacuum state~\cite{strombousso}.

Many interesting puzzles remain.  Most importantly we do not yet
have a controlled de Sitter background in string theory, or a
well-understood soliton whose near horizon limit leads to a
definition of de Sitter holography.   Indeed, it is not clear that
a stable de Sitter vacuum solution can be achieved.\footnote{The
papers of Hull~\cite{hull}, and Hull and Khuri~\cite{hullkhuri}
find de Sitter backgrounds in a variety of string theories with
unconventional signatures that are obtained via T-duality on
timelike directions.  While the stability of these theories
remains in question, they present an interesting potential
direction.  Another unconventional, but potentially unstable
approach appears in~\cite{andrewneil}.  It has been pointed out
that it might be possible to obtain de Sitter space from a
non-critical string theory~\cite{eva,alex}.  One approach to
Euclidean de Sitter space was suggested in~\cite{bhmds}.  Various
discussions of de Sitter space in the context of supergravity can
be found in \cite{nogo2, nogo1, nogo3, hullw, hullsolo, gates,
murat, nicolai, renata, bhmdswc,fre}.  Related developments in the the
study of time dependent backgrounds of string theory include
recent work on spacelike brane solutions~\cite{spacelike},
space-time orbifolds with fixed planes that are localized in
time~\cite{orbifold}, and Sen's proposal that dynamical rolling of
the tachyon of open string field theory can lead to interesting
consmologies~\cite{sen}.}

Even given a de Sitter background, the fact that the boundaries of
de Sitter space are euclidean surfaces at the beginning and end of
time completely changes the structure of potential holographic
bulk/boundary relationships.   Application of intuitions arising
from the AdS/CFT correspondence to de Sitter space also leads to
questions about the unitarity of possible dual field theories.

Given these puzzles,  we explore aspects of de Sitter physics that
are relevant to holography with a positive cosmological constant,
and uncover several new results.  In Sec. 2 we recall how the
AdS/CFT correspondence is formulated and explain why a naive
translation to de Sitter will not work.  We then display a
non-local map that commutes with the de Sitter isometries,
transforms the bulk-boundary propagator and solutions of free wave
equations in de Sitter onto the same quantitites in Euclidean AdS,
and takes the two boundaries of dS to the single EAdS boundary via
an antipodal identification reminiscent of the one advocated in
\cite{andyds}.  The existence of this map might suggest that at
the free field level some aspects of de Sitter physics could be
captured by the Euclidean field theory dual to Euclidean AdS.
However, the map has a non-trivial kernel, suggesting that not all
of de Sitter physics can be captured in this way.

Keeping the AdS/CFT correspondence in mind, we proceed to study
the action for scalar fields in de Sitter space as a functional of
boundary data.   To extend this investigation to gravity, we
display a family of solutions to 3d gravity with a positive
cosmological constant in which equal time sections are arbitrary
genus Riemann surfaces, and compute the action for these spaces as
a functional of boundary data in both the Einstein and
Chern-Simons formulations of 3d gravity.   In all these cases the boundary functional becomes trivial unless the contributions of data at each of the de Sitter boundaries is treated independently.

These studies, along with the non-local map discussed above,
suggest that if de Sitter space is dual to a Euclidean CFT, this
theory should involve two disjoint, but possibly entangled
factors.   We argue that these CFTs would be of a novel form, with unusual hermiticity conditions relating left movers and right movers. In Sec. 3 we explore  how such hermiticity conditions are implemented in a simple model with two chiral bosons.     We combine our observations to propose that a holographic dual description of de Sitter space would involve a pure entangled state in a product of two of our unconventional CFTs associated with the de Sitter boundaries.  In particular, we show that it is possible to construct such a state which is invariant under a single $SL(2,C)$ in a product of two theories that are separately $SL(2,C)$ invariant, as would be necessary for a holographic description of 3d de Sitter.   The decomposition of this state in a basis appropriate to antipodal inertial observers would lead to the thermal properties of the static patch.   In Sec. 4, we provide evidence for this picture by examining the physics of a scalar field in de Sitter space, explaining how the dS isometries are realized on it and how mode solutions in the global and static patches are related.  Sec. 5 concludes the paper by discussing open problems and future directions.

\section{Bulk/boundary maps and de Sitter holography}

In the spirit of the well known AdS/CFT correspondence~\cite{adsduality, adsduality2}, one could try to set up a correspondence between global de Sitter space and a Euclidean field theory on the de Sitter boundaries.   Specifically, the GKPW dictionary \cite{adsduality2} states that (for Euclidean AdS)
\begin{equation}
\langle \exp(\int_{bnd} \phi_0 O) \rangle  = Z_{bulk}(\phi_0)
\label{adsdualityform}
\end{equation}
where $Z$ is the bulk string partition function as a functional of boundary data (or $Z_{bulk}(\phi_0) = \exp(-S _{bulk}(\phi))$ in the supergravity limit), $\phi_0$ is the value of a bulk field $\phi$ at the spatial boundary of AdS, and $O$ is an operator in the dual CFT.  This precise formula also implies that the meaningful observables for string theory in AdS space are given by the correlation functions of the dual CFT.    In Lorentzian signature further subtleties are introduced in which non-fluctuating non-normalizable modes implement the boundary conditions on AdS and correspond to field theory sources, while  fluctuating normalizable modes in AdS space correspond to field theory states~\cite{lorentzian, lorentzian2}.\footnote{See \cite{braga} for another discussion in Poincare\'{e} coordinates for AdS.} The nature of the boundary CFT is determined by the fact that the near horizon geometry of certain solitonic states of string theory includes an AdS factor \cite{adsduality}.  In the case of de Sitter space, the absence of a direct argument for a duality from the physics of string solitons makes the discussion more tentative;  nevertheless, it is interesting to see what one gets.

Given the intuitions arising from the AdS/CFT correspondence and the fact that the early and late time infinities of de Sitter $\CI^\pm$ are natural holographic screens \cite{screens}, can one write a de Sitter analog of (\ref{adsdualityform})?  If so, one would attempt to write the following formula
\begin{equation}
\label{q1}
\langle \exp(\int_{I^{-}} \phi^{{\rm in}} O^{{\rm in}} + \int_{I^{+}}
\phi^{{\rm out}} O^{{\rm out}})
\rangle  = Z_{bulk}(\phi^{{\rm in}}, \phi^{{\rm out}})
\label{dsdef1}
\end{equation}
where $Z_{bulk}(\phi_{{\rm in}}, \phi_{{\rm out}})$ is the lorentzian bulk path integral for given the inital and final values of the bulk field $\phi$.

If we want to interpret the left hand side in (\ref{q1}) as the generating function of correlators  in some conventional euclidean CFT, equation (\ref{q1}) apparently involves two mismatched quantities. On one side we have a real quantity (given the real value of the generating functional) and on the other side we have a complex valued quantity - the lorentzian bulk path integral. The mismatch may point to a possible problem with unitarity in the dual CFT. Another interpetation of the mismatch is that the dual Hilbert space is equipped with a non-canonical Hilbert space structure.   We will find evidence for this in later sections.

This mismatch, however, is just one of the problems with specifying (\ref{q1}).   Recall that in the Lorentzian AdS/CFT correspondence, tbe boundary behaviour of fields is specified by the non-normalizable modes which grow near the AdS boundary.    Normalizable modes, which decay near the boundary, correspond to fluctuating states, and giving them a classical background value turns on VEVs for composite operators in the dual CFT \cite{lorentzian,lorentzian2}.  Fixing the non-normalizable mode corresponds to choosing a background, or, in the path integral language, picking boundary conditions for the path integral so that the fields approach a certain solution to classical equations of motion near the boundary.

By contrast, in de Sitter space there is no such distinction between normalizable and non-normalizable modes -- all mode solutions are normalizable.    There is a range of masses ($0 < 4 m^2 l^2< d^2$, in $d+1$ dimensional de Sitter, with cosmological scale $l$) in which a basis may be chosen in which the fields have a scaling growth or decay in time, but the solutions are still normalizable in the Klein-Gordon norm.   For larger masses we can pick a basis of modes that show oscillatory behaviour with a fixed frequency near infinity~\cite{andyds, strombousso, vacuum,volosprad}.  We will refer to these scaling/fixed frequency solutions at  early and late time infinity as $\phi^\pm_{{\rm in},{\rm out}}$.  All of this is reminiscent of the situation in AdS for masses in the range $- d^2  < 4 m^2 l^2 < 0$ (in $d+1$ dimensional AdS, with cosmological scale $l$  \cite{breitfreed,lorentzian}), where all mode solutions for scalar fields  are normalizable.   In these cases it is necessary to pick one scaling behaviour as dual to sources and the other as dual to VEVs \cite{lorentzian,lorentzian2}. While we might try the same strategy in de Sitter space, the fact that the classical equations of motion are second order implies that for classical solutions we can either specify both the ``source" and the ``VEV" scaling behaviour at one de Sitter boundary (e.g., $\phi^{\pm}_{{\rm in}}$), or we can specify the ``source" scaling behaviour at both de Sitter boundaries (e.g., $\phi^{+}_{{\rm in}, {\rm out}}$ as in (\ref{dsdef1})).   The latter description is better suited for a path integral setup as in (\ref{dsdef1}).  In effect, the sources at $\CI^+$ determine the VEVs at $\CI^-$ and vice-versa.

The authors of \cite{strombousso,volosprad} address the same issue by considering four CFT operators $O^\pm_{{\rm in}, {\rm out}}$ that are related to each of the boundary behaviours described above.  In particular~\cite{volosprad} proposes a method for calculating the correlation functions of such CFT operators from a de Sitter analog of the boundary S-matrix calculations of AdS space \cite{adssmat}.    If there is a bulk-boundary relationship in de Sitter space there can only be two independent operators of this kind, so, as discussed in~\cite{volosprad} there is a relation between $O^\pm_{{\rm in}, {\rm out}}$ on shell.    Consequently, it is not entirely clear how the generating functional of the correlators of these operators computed by the prescription in~\cite{volosprad} can be related to a natural de Sitter bulk quantity like the path integral.

Another outstanding issue  is whether the left hand side of (\ref{dsdef1}) should be regarded as a theory that lives on a single sphere, or to a theory that lives on two separate spheres.  In \cite{andyds} it was argued that the theory dual to de Sitter should live on a single sphere in view of the antipodal singularities of propagators in de Sitter space.  Certainly, for free fields, the values of the fields at $I^-$ are directly expressed in terms of the values of the fields at $I^+$, and it seems reasonable to view everything as living on a single sphere. However, once we turn on interaction, and turn to the full quantum gravitational path integral, the relation between the two boundaries becomes much less manifest. For instance, one can put two different metrics on the two boundaries. We will indeed find indications that the dual theory lives on two separate spheres.

In \cite{wittds} Witten suggested that the bulk lorentzian path integral could be naturally associated to an inner product of two states in a complex Hilbert space of the dual theory (given a suitable action $\Theta$ of the CPT symmetry group). Given the bulk path integral
\be \label{q3}
\langle f|i \rangle = Z_{bulk}(\phi_{i}, \phi_{f})
\ee
the inner product $(,)$ is defined via $(f,i) = \langle \Theta f|i \rangle$.  (The CPT transformation here maps the "out" states into "in" states allowing us to construct an inner product for the latter.) This suggests
that perhaps a de Sitter holographic dictionary (if it exists) would be defining a Hilbert space structure and not an S-matrix, although \cite{wittds} does propose a method to relate the de Sitter lorentzian path integral to local correlation functions of a dual field theory.  Still, the precise definition of the right hand side of (\ref{q3}) runs into the same problems as we discussed above.

Given these different interpretations and possibilities, in the remainder of this section we explore aspects of the physics of de Sitter that are relevant for holography, either by making sense of (\ref{dsdef1}) or by a suitable modification of this equation.   Specifically, we  review harmonic analysis on de Sitter space, and point out that there is an interesting non-local map from Lorentzian de Sitter space to Euclidean anti-de Sitter space. This map commutes with the action of the isometry group, and therefore maps solutions of free field equations into solutions. Such a non-local map suggests the interesting possibility of understanding de Sitter holography using the  well known bulk/boundary dictionary from the AdS/CFT duality.   Next, we discuss the on-shell action for free scalar fields, and point out the problems with the application of the standard GKPW-like machinery~\cite{adsduality2}  to de Sitter space. After that, we turn to gravity in $2+1$ dimensions with a positive cosmological constant and display a family of solutions in which equal time sections are arbitrary genus Riemann surfaces.  We consider the on-shell action of these spaces  and again find subtleties with the standard GKPW approach.   For example, the conformal anomaly contributions from the surfaces at $\CI^\pm$ cancel against each other.  Finally, we discuss some aspects of the $2+1$ dimensional case from the Chern-Simons theory perspective.   A basic conclusion from these studies is that if there is a dual to de Sitter, it probably lives on two spheres rather than one, and that a GKPW-like formula such as (\ref{dsdef1}) cannot be naively applied.

\subsection{A non-local map from dS to AdS}

To gain insight into holography in de Sitter space it is interesting to ask whether there is a map from the physics of Lorentzian de Sitter to Euclidean AdS in view of (\ref{hyperboloid}).   Such a map cannot amount to mere analytic continuation since the scalar Green function on AdS does not continue to a 2-point function in a de Sitter vacuum \cite{strombousso}.  Also, since de Sitter space has two boundaries while Euclidean de Sitter has only one both dS boundaries must map into the single AdS one.   As a preliminary step we show below that there is an interesting nonlocal map that maps solutions to free wave equations in dS and EAdS onto each other while preserving the isometries of the hyperboloid (\ref{hyperboloid}).   Note that the non-locality implies that this map cannot be used to transform the local physics of AdS or its dual into the local physics of dS or its dual.

Some of  the analysis below is carried out in three dimensions since it is convenient since to use the representation theory of SL(2,C); higher dimensional extensions of the results should be straightforward.

\subsubsection{Harmonic Analysis}

The harmonic analysis of $dS_3$ is discussed in chapter 6, section~4 of
\cite{gelfand5}. There, the decomposition of $L^2(dS_3)$ in terms
of irreducible representations of $SL(2,C)$ is given. To describe this
result, we start with the
description of lorentzian de Sitter space as
a hyperboloid immersed in the flat Minkowski space.
de Sitter space is described by
\be \label{exx1}
P(X,X)=1
\ee
where
\be \label{defP}
P(X,Y) = -X_0 Y_0 + X_1 Y_1 + \ldots + X_d Y_d
\ee
represents a Minkowski signature quadratic form (we have set the cosmological constant to 1).
The decomposition of a function $f(X)$ in $L^2(dS_3)$ reads
\bea
f(X) &  = & \frac{1}{2 (4\pi)^3} \int_{0}^{\infty} \rho^2 d\rho
\int_{S^2} F(U(\omega);\rho) |P(X,U(\omega))|^{-\frac{1}{2} i\rho-1}
d\omega \nonumber \\
& & + \frac{4}{\pi^2} \sum_{n=1}^{\infty} n \int_{S^2}
F(U(\omega),Y;2n) e^{2in \theta(Y,X)} \delta( P(U(\omega),X)) d\omega
\label{fourier}
\eea
where $\cos\theta(Y,X)=-P(X,Y)$, $Y$ is any point on $dS_3$,
and $U(\omega)$ is a family of points on the null cone $P(U,U)=0$ which
are parametrized by $\omega \in S^2$. If we write $\omega$ as
$(U_1,U_2,U_3)$ with $U_1^2+U_2^2 + U_3^2=1$, then $U(\omega)=
(1,U_1,U_2,U_3)$. Thus $U(\omega)$ sweeps out a two-sphere on
the forward null cone. Finally, $d\omega$ is the standard measure
on the two-sphere. The two-sphere can be replaced by any other
two-surface homeomorphic to it with a suitable change in the
measure $d\omega$.   The Fourier modes labeled by $F$ can be obtained from $f(X)$ via
suitable integrals of $f(X)$ and are independent of $Y$, see \cite{gelfand5} for details.

The first term in (\ref{fourier}) contains principal series representations of
$SL(2,C)$, similar to what one finds in the case of euclidean AdS. In the
latter case, one typically analytically continues the functions
$|P(X,U(\omega))|^{-\frac{1}{2} i\rho-1}$ to imaginary values of
$\rho$, in which case these functions become the wave functions discussed
in e.g. \cite{teschner9712,boer9812}. At the same time, they are
are also equal to the bulk-boundary propagator introduced by Witten
\cite{adsduality2}.

To make the last statement explicit, we express $|P(X,U)|$ in terms
of coordinates on the inflationary patch.  In Euclidean AdS, we
parametrize $X$ as
\be  \label{ff2}
X=(\frac{1}{2} (e^t + e^{-t} + e^t z_i z^i),
\frac{1}{2} (e^t - e^{-t} - e^t z_i z^i),e^t z_1, \ldots,e^t z_{d-1})
\label{Xparam}
\ee
from which we obtain the metric $ds_{{\rm EAdS}}^2 = dt^2 + e^{2t} (dz_0^2 + \cdots + dz_{d-1}^2)$ with $t$ as the radial direction, $z_0$ as Euclidean time and the AdS boundary at $t \rightarrow \infty$, in which limit the  EAdS hyperboloid approaches the null surface $P(U,U) = 0$. .   The conformal boundary of EAdS is therefore a sphere embedded in this null  asymptote.   Conformal rescaling of the boundary sphere moves it along the null cone $P(U,U) = 0$.   Therefore writing  points on the (conformally rescaled) boundary sphere as
\be
U=(\frac{1}{2} (1 + y_i y^i),
\frac{1}{2} (1 - y_i y^i),y_1, \ldots,y_{d-1}) ,
\label{Uparam}
\ee
we obtain
\be \label{bb1} |P(X,U)|_{\rm EAdS} = \frac{1}{2} (e^{-t} + e^t (y-z)^2) .
\ee
Raising this to the  suitable power indeed gives the bulk-boundary
propagator of euclidean AdS. To get Lorentzian de Sitter space, we replace the term
$e^{-t}$  by $-e^{-t}$ in $X^0$ and $X^1$ in the parametrization (\ref{Xparam}). This
gives the metric  $ds^2 = -dt^2 + e^{2t} d\vec{z}^2$ which covers half of de Sitter (the inflating patch).  As $t \rightarrow \infty$ the $\CI^+$ the de Sitter hyperboloid approaches to the null surface $P(U,U) = 0$.   The conformal boundary of de Sitter is therefore a sphere embedded in the cone $P(U,U) = 0$,  just like the EAdS boundary parameterized in (\ref{Uparam}).   Conformal rescaling of the boundary amounts to rescaling $U$.  With these parametrizations we find that
\be \label{bb2} |P(X,U)|_{dS} = \frac{1}{2} |-e^{-t} + e^t (y-z)^2| .
\ee
Again, raising this to the power in (\ref{fourier}) gives the bulk-boundary propagator in de Sitter space (see \cite{volosprad} for a discussion using conformal time for the inflationary patch).

Actually, the bulk-boundary propagator $|P(X,U)|^{-1-i\rho/2}$ is not well-defined as it stands, because
$P(X,U)$ changes sign as a function of $t$. We can define it once
we choose a branch cut above or below the real $t$-axis
in the complex $t$-plane. If we choose the branch cut above the $t$-axis,
the functions $|P(X,U)|^{-1-i\rho/2}$ are like the Euclidean modes
in de Sitter space, because they are analytic on the lower hemisphere
of the Euclidean sphere that is obtained by the analytic continuation of
dS space.

It is straightforward to express everything in global coordinates for de Sitter with metric $ds^2 = -dt^2 + \cosh^2 t \, d\Omega^2$.
We can take e.g. $X=(\sinh t, {\bf n} \cosh t)$ and $U=(1,{\bf
m})$ with ${\bf m,n}$ unit vectors in ${\bf R}^d$. Then
 \be
\label{bb3} |P(X,U)|_{dS} = |-\sinh t + ({\bf n} \cdot
{\bf m}) \cosh t | . \ee

The second term in (\ref{fourier}) is interesting because it does
not have a counterpart in AdS. The Fourier components
$F(U(\omega),Y;2n)$ are obtained from the integral of $f(X)$ along
null geodesics. A null geodesic in de Sitter space is given by a
straight line $Y+tU$, where $Y$ is a point on de Sitter space, so
that $P(Y,Y)=1$, and $U$ is a point on the cone $P(U,U)=0$ that
satisfies $P(Y,U)=0$.

In summary, the decomposition of $L^2(dS_3)$ is quite similar to
the decomposition of $L^2(EAdS_3)$, except that $L^2(dS_3)$
contains an extra set of representations associated to the
integrals of functions along null geodesics.

\subsubsection{A non-local map from dS to AdS}

It has been suggested that quantum gravity on de Sitter space is
dual to a single Euclidean CFT \cite{andyds}. On the other
hand, in the standard AdS/CFT correspondence, Euclidean CFT's are
dual to quantum gravity on Euclidean Anti-de Sitter space. Thus
it is worth exploring whether there exists a relation between de Sitter
space and Anti-de Sitter space that is more subtle than
analytic continuation. To find such a relation, we will now look
for a (possibly non-local) map from de Sitter space to anti-de Sitter space
that commutes with the isometry group $SO(d,1)$ that de Sitter and
Euclidean Anti-de Sitter have in common.

We consider a non-local map to of the form
  \be \psi(Y) = \int dX
K(Y,X) \phi(X) \ee where $dX$ denotes the invariant measure on
$dS_d$, and demand that the kernel $K(X,Y)$ has the fundamental property that it commutes
with the $SO(d,1)$ actions. In other words,
\be K(gX,gY)=K(X,Y) \ee
for $g\in SO(d,1)$.

Using suitable $g$, we can always achieve that
$X_i=0$ for $i<d$, and $X_d=1$. Call this point $E$.
Then it is sufficient to know $K(E,Y)$, because
by acting with the group we can recover the rest of
$K$ from this.

The point $E$ is preserved by a $SO(d-1,1)$ subgroup. This can be
used to put $Y$ in the form \be Y(\xi)\equiv (Y_0,\ldots,
Y_d)=(\sqrt{1+\xi^2},0,\ldots,0,\xi) . \ee The group action cannot
be used to change the value of $\xi$. Thus, the kernel is
completely determined by a function of a single variable, \be
K(E,Y(\xi))\equiv K(\xi). \ee It is easy to write down the kernel
explicitly, once we are given $K(\xi)$. Clearly $P(gX,gY) =
P(X,Y)$, and we also observe that $P(E, Y(\xi)) = \xi$. Therefore,
\be \label{kernel} K(X,Y) = \int d\xi \delta(P(X,Y) - \xi) K(\xi)
\ee satisfies all the required properties, it is group invariant
and reduces to $K(\xi)$ for $X=E,Y=Y(\xi)$. The simplest form of the kernel is $K(\xi)=\delta(\xi)$. In this
case our transform is closely related to the Radon transform
\cite{radon} known in the study of tomography.

We now investigate the transform for a free massive scalar.
In a slightly different notation from the one before,
we label points on the $d-1$ sphere by unit normals ${\bf n,m}$.
In general the transform is
\be
\psi(\rho,{\bf m}) = \int d\xi \int dt d^{d-1} {\bf n} \cosh^{d-1} t
K(\xi) \delta(-\sinh t \cosh \rho + \cosh t \sinh \rho ({\bf n} \cdot
 {\bf m}) -\xi) \phi(t,{\bf n}) .
\ee
Here, we parametrized a point on dS as $X=(\sinh t,
{\bf n} \, \cosh t)$ and a point on $EAdS$ as $Y=(\cosh \rho, {\bf
m} \, \sinh \rho)$ ($\rho \geq 0$) giving the metrics $ds^2_{{\rm dS}} = -dt^2 + \cosh^2 t \, d\Omega^2$ and $ds^2_{{\rm EAdS}} = d\rho^2 + \sinh^2\rho \, d\Omega^2$.   The dS boundaries are at $t \rightarrow \pm \infty$ and the EAdS boundary is at $\rho \rightarrow \infty$.

It is not so easy to do this integral for a global mode solution,
because we have to integrate spherical harmonics and
hypergeometric functions, and the result is a complicated linear
superposition of modes in EAdS. To get some idea of what this
transform does, we apply it to a Green's function $G(t,{\bf
n},t_0,{\bf n}_0)$ that obeys \be \triangle G =
\delta(t-t_0)\delta({\bf n}-{\bf n}_0). \ee The transform of $G$
obeys \be \triangle \psi = K(-\sinh t_0 \cosh \rho + \cosh t_0
\sinh \rho ({\bf n}_0
 \cdot {\bf m}) ) .
\ee
Thus, even though we start with a localized source for the Green's
function, after the transform we find a source that is smeared out
over all of $EAdS$.

In case $K(\xi)=\delta(\xi)$, we get
\be
\triangle \psi = \delta(-\sinh t_0 \cosh \rho + \cosh t_0 \sinh \rho \,  ({\bf n}_0
 \cdot {\bf m}) ).
\ee
The right hand side contributes on the surface
\begin{equation}
 \label{j8}
\tanh \rho = \frac{\tanh t_0}{({\bf n}_0 \cdot {\bf m} )}
~~~;~~~ {\rm sign}(t_0) = {\rm sign}({\bf n}_0 \cdot {\bf m})
.
\end{equation}
The condition on the sign arises because $\rho \geq 0$ and we see that antipodal points in de Sitter ($(t_0, {\bf n}_0)$ and -$(t_0, {\bf n}_0)$) get mapped to the same points in Euclidean AdS $(\rho,{\bf m})$ that solves (\ref{j8}).   If $t_0>0$, $\rho$ takes it smallest value at $\rho=t_0$, while for  $t_0<0$, $\rho$ has as smallest value $|t_0|$, when ${\bf m}=-{\bf n}_0$.  The de Sitter boundaries at $t \rightarrow \pm \infty$ are mapped into the EAdS boundary at $\rho \rightarrow \infty$.

Although the map is non-local, equation (\ref{j8}) shows that the
support of $\triangle \psi$ is on a surface with $\rho \leq
|t_0|$. Thus, once we send $|t_0|$ off to infinity, $\rho$ should
go to infinity as well.   Therefore we expect that the bulk-boundary propagator should have a simple transform.  As an example we work out the $d=3$ case where this propagator is given by
 $|P(X,U)|^{-1-i\rho/2}$.   Here $X$ is a point in de Sitter space ($P(X,X) = 1$) and $U$ is a point in the conformal boundary of de Sitter space specified by $P(U,U) = 0$ as discussed earlier.    Hence, in transforming from de Sitter to EAdS, we map $X$ into points $Y$ in EAdS, but $U$ is simply a parameter.
The  transform (with $K(\xi) = \delta(\xi)$) is given by
\be \label{j88}
\psi(Y,U) = \int \delta (P(X,Y))|P(X,U)|^{-1-i\rho/2} dX
\ee
with $Y\in EAdS$ and $U$ a point on the cone $P(U,U)=0$.    Since the conformal boundary of EAdS also lies in the cone $P(U,U)$, we can treat $U$ equally as parametrizing boundary points in EAdS or de Sitter (the latter subject to an antipodal identification as described above).    Since
$\psi(Y,U)$ satisfies $\psi(Y,U)=\psi(gY,gU)$ for $g \in SO(3,1)$
we can without loss of generality take $Y=(1,0,0,0)$ and
$U=(\xi,\xi,0,0)$. If we parametrize $X=(\sinh t, \cosh t {\bf
n})$ with ${\bf n}=(n_1,n_2,n_3)$ a point on the unit two-sphere,
the right hand side of (\ref{j88}) becomes
\be
\int dt d^2{\bf n} \cosh^2 t \delta(-\sinh t) |-\xi \sinh t + \xi
\cosh t n_1|^{-1-i\rho/2} .
\ee
This is proportional to $|\xi|^{-1-i\rho/2}$. As $\xi$ can be
invariantly written as $P(Y,U)$, we conclude that $\phi(Y,U)$ is
proportional to $P(Y,U)^{-1-i\rho/2}$. Therefore, the non-local
map takes the bulk-boundary propagator of dS into the bulk-boundary
propagator of EAdS.   Since solutions to the free wave equations in EAdS and in de Sitter can be constructed by convolving the bulk-boundary propagator with boundary data, we see that such solutions map onto each other.

Note that in the AdS/CFT correspondence the parameter $i\rho$ here is related to to the conformal dimension of the dual operator and real (imaginary) $\rho$ corresponds to an imaginary (real) conformal dimension associated with fields below (above) the Breitenlohner-Freedman mass bound for AdS \cite{breitfreed}.   In de Sitter space it has been noted that there is a sort of inversion of the Breitenlohner-Freedman bound -- fields with masses above a certain bound would map onto operators with imaginary conformal dimension in a dual CFT while lower masses would map to operators with real dimension \cite{andyds}.   Here the reality of the conformal dimension is being preserved in the dS-EAdS map.   If we want to have real dimensions on both sides of the map we must stick to fields in de Sitter with masses lower than the the bound in \cite{andyds}.   However, as we will discuss later in this paper there may be ways of interpreting apparent imaginary conformal dimensions in CFTs with different hermiticity conditions from the usual ones.

Altogether we have constructed a non-local map from dS to EAdS
that commutes with the isometries, maps the two boundaries of dS
onto the single boundary of EAdS, using an anti-podal
identification for one of the two, and maps bulk-boundary
propagators to bulk-boundary propagators. It therefore is tempting
to define a dS/CFT correspondence by first mapping everything to
EAdS, and  subsequently applying the standard rules of AdS/CFT.
This certainly yields a single Euclidean CFT. However, since the
map is non-local, it does not map local interactions of fields to
local interactions of fields, and does not naively extend to a
well-defined map on the level of (super)gravity. More importantly,
the map has a kernel \cite{gelfand5}, which is similar to the
second line in (\ref{fourier}). If we know the field on EAdS, we
cannot reconstruct the integrals of the field on dS along null
geodesics. That information is lost under this map. We view this
as suggestive evidence that it is not sufficient to describe dS space in
terms of a single CFT. We will later advocate a picture where the
dual of dS involves two (entangled) conformal field theories.

It would be interesting further  to  study this non-local map, and to find
out whether it allows us to set up a GKPW-like formulation for de Sitter space, and to extend this map to other time-dependent backgrounds.

\subsection{On-shell action-scalar fields}

To make sense of expressions of the form (\ref{q1}) we need to define the left hand side via the right hand side.  Naively, as in
AdS/CFT, this is the on-shell value of the (super)gravity action
evaluated with given boundary conditions. At tree-level, this
becomes the (super)gravity action evaluated on a solution of the
equations of motion.
For a free scalar field with action
\be \label{qa1}
S= \int_M d^dx \sqrt{-g} (g^{\mu\nu} \partial_{\mu} \phi
\partial_{\nu} \phi + m^2 \phi^2)
\ee
the action reduces for a solution of the equations of motion to
\be \label{qa2}
S=-\int_{\partial M} d^{d-1}x \sqrt{\hat{g}} \, \phi \, \partial_t\phi =
- \int_{\CI^+}  \sqrt{\hat{g}} \, \phi  \, \partial_t\phi +
\int_{\CI^-} \sqrt{\hat{g}} \,  \phi \,  \partial_t\phi
\ee
when we choose a de Sitter metric of the form
\be
\label{qa3} ds^2 = -dt^2 + \hat{g}_{ij} dx^i dx^j .
\ee
As discussed earlier, all solutions to the scalar field equations are normalizable, hence in order to impose boundary conditions in the path integral in might be necessary to add boundary terms to the actions
in (\ref{qa1}) and (\ref{qa2}).
It is easy to show that
as $t\rightarrow -\infty$, solutions to the field equations behave as \cite{vacuum,strombousso,volosprad}
\be \label{qa4}
\phi \sim \phi_{\rm in}^+ e^{2h^+ t} + \phi_{\rm in}^- e^{2 h^- t}
\ee
and as $t\rightarrow +\infty$ as
\be \label{qa5}
\phi \sim \phi_{\rm out}^+ e^{-2h^+ t} + \phi_{\rm out}^- e^{-2
h^- t} .
\ee
The scaling dimensions are
\be
2 h^{\pm} =\frac{d-1}{2} \pm \sqrt{ \frac{(d-1)^2}{4} - m^2}
\label{scaldim}
\ee
and we will assume that we are in the regime $4m^2>(d-1)^2$ so
that $h^{\pm}$ are complex\footnote{For $4m^2 <(d-1)^2$ the
weights $h^{\pm}$ are real and the situation in de Sitter space is
more closely related to the situation in Euclidean AdS with
growing and decaying modes at the boundaries. Nevertheless, many
of the problems we discuss here are not resolved by restricting
attention to real $h^{\pm}$ only.   Likewise, as discussed in the previous section, the solutions to the wave equation are normalizable for any value of $m^2$.}. In (\ref{qa2}) there are
contributions $\phi^+ \dot{\phi}^+$, $\phi^+ \dot{\phi}^-$,
$\phi^- \dot{\phi}^+$ and $\phi^- \dot{\phi}^-$ at each boundary.
The first and the last of these scale as $\exp(\pm 4 h^{\pm} t)$.

In standard AdS/CFT correspondence we similarly get a $\phi \, \partial_\rho \phi$ boundary action (where $\rho$ is the radial direction away from the boundary).   We rewrite this as a nonlocal expression in terms of boundary data by using a bulk-boundary propagator, and regulate  by adding local counterterms to remove divergences.   For a scalar field such counterterms can be built out of the boundary value of the field and its covariant derivatives with respect to the boundary coordinates.   The surviving finite terms yield CFT correlation functions.  The multi-local terms give n-point functions at separated points and the local parts are associated with contact terms and one-point functions.   In particular, in Lorentzian signature,  local terms arising from $\phi^+\phi^-$ type terms in the language used above lead to one-point functions in a CFT state determined by the $\phi^+$ scaling piece of the bulk solution.  In de Sitter space, with $4m^2 > (d-1)^2$, this strategy cannot be applied directly.   In (\ref{qa4},\ref{qa5}) we have only specified the leading scaling behaviour of mode solutions.  After accounting for the boundary metric, the leading and subleading terms in the $\phi^+ \dot\phi^+$ and $\phi^- \dot\phi^-$ contributions to the boundary action, as well as in possible local counterterms, scale as $\exp(\pm (4 h^{\pm} - 2 n)t)$ for integer  $n$.   Since the square root in (\ref{scaldim}) is imaginary this means that that there are no terms arising from $\phi^{\pm}\dot\phi^\pm$ contributions that approach a finite and well-defined limit as they approach the de Sitter boundary -- they either oscillate or vanish after regulation.   In conformal time as opposed to the global time used above the oscillations will be wild, growing faster as the boundary is approached.  It seems natural to remove any such wildly oscillating behavior, in which case nothing is left.  (When $4m^2 < (d-1)^2$ these oscillations do not arise, but other issues are relevant in that case.)

The terms with a finite limit as we approach the boundary are
$\phi^+ \dot{\phi}^-$ and $\phi^- \dot{\phi}^+$. In the Euclidean AdS/CFT
case, $\phi^+$ and $\phi^-$ are not independent, but one is given
in terms of the other by requiring that the solution is regular in
the interior of the Euclidean AdS space, while in the Lorentzian case $\phi^+$ is indepdendent and related to the specification of a CFT state.  In either case, we obtain
non-trivial finite contributions from this term, related to contact terms (which we may wish to remove) or to one-point functions in a CFT state.  In the case of  Lorentzian de
Sitter space, $\phi^+$ and $\phi^-$ are independent. Actually,
they are each others complex conjugate, and it is really the real
and imaginary part of $\phi^+$ that are independent. The on-shell
action becomes
\be \label{qa8}
S = (d-1) \int_{I^-} d^{d-1}x \sqrt{\hat{g}} \phi_{\rm in}^+ \phi_{\rm in}^- -
 (d-1) \int_{I^+} d^{d-1}x \sqrt{\hat{g}} \phi_{\rm out}^+ \phi_{\rm out}^-
\ee
where the factor of $d-1$ arises from the time derivatives,
$d-1=2(h^+ + h^-)$.   Earlier we discussed that  we might follow the AdS/CFT philosophy and associate operators and VEVS with $\phi^\mp$ scalings or, alternatively, following \cite{strombousso,volosprad} we could try to associate differnt CFT operators $O^\mp $ with these scalings.  In these two cases (\ref{qa8}) would be associated with VEVs of CFT operators or with contact terms between operators respectively.
 In addition, the two terms in (\ref{qa8}) are not
independent, but are proportional to each other, because
$\phi_{\rm out}^{\pm}$ are linear combinations of $\phi_{\rm
in}^{\pm}$ for a solution of the classical equations of motion.

Thus, the naive on-shell action for a free scalar field merely gives
a contact term. It is an interesting question whether there are other
ways to define the supergravity action, which yields results that
resemble the AdS/CFT answer and produces CFT two-point functions. One direction
is to use the bulk-boundary propagator and to add suitable non-local
boundary terms to the action, for instance as in \cite{shomer0203168}.  Normally, we do not add such a non-local quantitites -- while local counterterms may be used to remove divergences in an effective action, non-local terms modify the correlation functions.   However, nonlocal boundary terms may be acceptable if they arise as a method of imposing boundary conditions in de Sitter space that are necessary for the validity of a correspondence with a dual CFT.    We seek to impose boundary conditions on a free scalar field that lead to an non-trivial action for solutions to the equations of motion.   Given the asymptotic behavior of the fields,
\be
\phi \sim \phi_{\rm in}^+ e^{2h^+ t} + \phi_{\rm in}^- e^{2 h^- t}
\ee
one notices that
\be
\lim_{t\rightarrow -\infty} e^{-2 h^+ t} (\partial_t
 - 2 h^-) \phi = 2(h^+-h^-) \phi_{\rm in}^+ + {\rm subleading} .
\ee
Therefore, fixing the left hand side of this equation imposes
a suitable boundary condition on $\phi$. At the same time, a
general solution of the equations of motion is of the form
\be \label{qa11}
\phi(t,{\bf m}) = \int d{\bf n} \rho({\bf n})
|-\sinh t + \cosh t ({\bf n}\cdot {\bf m})|^{-2 h^+}.
\ee
If we define ${\bf A}(\rho)$ via
\be
{\bf A}(\rho) = e^{-2 h^+ t} (\partial_t
 - 2 h^-)\int d{\bf n} \rho({\bf n})
|-\sinh t + \cosh t ({\bf n}\cdot {\bf m})|^{-2 h^+}
\ee
then the idea, following \cite{shomer0203168} is to add a boundary term that imposes the boundary
condition
\be
e^{-2 h^+ t} (\partial_t
 - 2 h^-) \phi = {\bf A}(\rho) .
\ee
The relevant non-local boundary term is proportional to
$\int \phi \, A(\rho)$, and when evaluated on clasical solutions
of the equation of motion yields the CFT two-point function for
$\phi^+$, at least in the case of the AdS/CFT correspondence.
In the dS/CFT case, there are some extra complications. First,
one has to be quite careful in defining integrals like
(\ref{qa11}), which have a singularity, and second, one has
to impose the boundary conditions on both sides simultaneously.
We have not studied this in detail, but it would be nice to
understand this better.

Yet another approach to define correlation functions from supergravity
is the S-matrix inspired approach of \cite{andyds,strombousso,volosprad}. In this approach,
each of the scaling behaviours $\phi_{\rm in}^{\pm}$ of bulk fields is associated with the insertion of a CFT operator on $\CI^+$ or $\CI^-$, and a method inspired by by the boundary S-matrix for AdS space was proposed for calculating the correlation functions of operators.  CFT-like two point functions and contact terms are obtain.  However, unlike the AdS case, it is not clear how such de Sitter calculations are related to a bulk quantity with a classical limit like the path integral.   In particular since $\phi_{\rm in}^{\pm}$ are not all independent of each other classically, an arbitrary correlation function of corresponding operators $O^{\pm}_{\rm in,out}$ cannot be computed from any quantity that has a classical limit in  the spacetime theory.  Indeed in the free field theory,  if we tried to associate the correlators of $O^{\pm}_{\rm in,out}$ with a bulk path integral as
\begin{equation}
 \label{q2}
\langle \exp(\int_{I^{-}} \phi_{\pm}^{{\rm in}} O_{\mp}^{{\rm in}} + \int_{I^{+}}
\phi_{\pm}^{{\rm out}} O_{\mp}^{{\rm out}})
\rangle  = Z_{bulk}(\phi_{\pm}^{{\rm in}}, \phi_{\pm}^{{\rm out}}) .
\end{equation}
then the computation on the right hand side is dominated by a saddlepoint only when we restrict $\phi$ to solutions to the equations of motion.  In this case, the proposed CFT operators $O^{\pm}_{\rm in,out}$ that couple to  $\phi_{\rm in}^{\pm}$  satisfy, ithe property that
\be
\phi^+_{\rm in} O^-_{\rm in} + \phi^-_{\rm in} O^+_{\rm in} =
\phi^+_{\rm out} O^-_{\rm out}   + \phi^-_{\rm out} O^+_{\rm out}
\ee
if $\phi$ is a solution of the equation of motion.   Then if as in \cite{andyds,strombousso,volosprad} we associate all of these operators with a single sphere,  the answer becomes $Z=1$.
If the operators are associated with  different spheres, the answer becomes more interesting.

In the next section, we will see that the above observations also apply to some
extent once we look at the Einstein-Hilbert part of the action. At this point,
there could be two points of view. One could take the point of view that
the naive supergravity action which yields just a contact term is the
right answer. This would be very close to the proposal of Witten \cite{wittds}
because the contact term is closely related to the inner product in the free
field theory on de Sitter space, and free field action seems to compute
nothing but this inner product. In this case, all the interesting physics
would be hidden in interactions and in higher order effects. On the other
hand, a much richer structure is obtained using the S-matrix approach of
\cite{andyds,strombousso,volosprad}. As we will explain later, these two points of view do
not necessarily have to disagree with each other. Our final proposal
for a dual description of de Sitter space will involve two CFT's
living on two different boundaries. These two theories are entangled.
From the S-matrix point of view, we are computing correlation functions
in these entangled CFT's. However, one might equally well argue that
the set of correlation functions in one of the two CFT's represent
the wave functions of the Hilbert space of the theory (very much as
in the relation between Cherns-Simons theory and CFT), and the
correlation functions in the entangled state represent the inner
product on this Hilbert space. Thus, our final proposal incorporates
both points of view in a satisfactory manner, though in order to probe
the full structure the S-matrix point of view is needed.

\subsection{On-shell action-gravity}

In this subsection we investigate the gravitational
on-shell bulk action of de Sitter space.
We start by describing ``new'' classical solutions
of 2+1 dimensional gravity with a positive cosmological constant.
The solutions we find are very similar to the solutions
of 2+1 gravity with a negative cosmological constant, that
were found in the Fefferman-Graham approach in the context of the
AdS/CFT duality, see
e.g. \cite{ss}, \cite{rooman} and \cite{krasnov}.\footnote{Other new solutions of 3d gravity with a positive cosmological constant appear in \cite{jens}.  It would be interesting to find the relation of these solutions with ours.} 

The most general solution reads
\bea
ds^2_{I^+} & = & -dt^2 + e^{2t} e^{\phi} dw d\bar{w} -\frac{1}{2}
 (T^{\phi}-T(w))dw^2 -\frac{1}{2} (\bar{T}^{\phi}  - \bar{T}(\bar{w}))
 d\bar{w}^2 -R dw d\bar{w} \nonumber \\
& & +\frac{1}{4} e^{-2 t} e^{-\phi} ((T^{\phi}-T(w))
dw + Rd\bar{w})((\bar{T}^{\phi}  - \bar{T}(\bar{w})) d\bar{w} + R dw),
\eea
where
\be
T^{\phi} = \partial_w^2 \phi - \frac{1}{2}(\partial_w \phi)^2,
\quad
\bar{T}^{\phi} = \bar{\partial}_{\bar{w}}^2 \phi -
\frac{1}{2} (\bar{\partial}_{\bar w} \phi)^2 ,
\quad
R=\partial_{w} \bar{\partial}_{\bar{w}} \phi.
\ee
Here $T(w)$ and $\bar{T}(\bar{w})$ are (anti)holomorhic
quadratic differentials.
The metric at $I^+$ is conformal, $ds^2=e^{\phi} dw d\bar{w}$;
Up to the finite number of free parameters contained in $T(w),\bar{T}(\bar{w})$
the metric at $I^-$ is completely determined in terms of that on $I^+$
as
\be ds^2_{I^-} = \frac{1}{4} e^{-\phi} ((T^{\phi}-T(w))
dw + Rd\bar{w})((\bar{T}^{\phi}  - \bar{T}(\bar{w})) d\bar{w} + R dw) .
\ee
This is presumably a feature of $d=2+1$ dimensions where the metric
has no dynamical degrees of freedom. In higher dimensions one would
expect to be able to independently fix the metric at $I^{+}$ and $I^-$.

We can get any Riemann surface of genus $g>1$ by taking $w,\bar{w}$
to live in the complex upper half plane and by modding out
by a discrete subgroup $\Gamma \subset PSL(2,{\bf Z})$.
This will work if the holomorphic quadratic differential
is invariant under $\gamma\in\Gamma$,
\be
T(\gamma(w)) \left(\frac{\partial{\gamma}}{\partial w} \right)^2 =
T(w)
\ee
and
\be
\phi(\gamma(w)) = \phi(w) - \log \left|\frac{\partial{\gamma}}{\partial w}
\right|^2 .
\ee

It is interesting to compare this to the most general solution
with negative cosmological constant, which reads (in Euclidean
space)
\bea
ds^2 & = & dt^2 + e^{2t} e^{\phi} dw d\bar{w} +\frac{1}{2}
 (T^{\phi}-T(w))dw^2 +\frac{1}{2} (\bar{T}^{\phi}  - \bar{T}(\bar{w}))
 d\bar{w}^2 +R dw d\bar{w} \nonumber \\
& & +\frac{1}{4} e^{-2 t} e^{-\phi} ((T^{\phi}-T(w))
dw + Rd\bar{w})((\bar{T}^{\phi}  - \bar{T}(\bar{w})) d\bar{w} + R dw).
\eea
Note that this clarifies the relation between the Fefferman-Graham (FG) expansion
in $\Lambda<0$ and $\Lambda>0$ \cite{mott, fefgrah}. Given a solution to the
Euclidean equations of motion, given in an FG expansion, we
first send $e^{2t}\rightarrow e^{-2t}$, in other words we
shift $t \rightarrow t+\pi i /2$. The metric will still solve
the $\Lambda<0$ equations of motion but no longer be Euclidean.
Next we send $g_{\mu\nu} \rightarrow -g_{\mu\nu}$. This gives a Minkowski
solution of the $\Lambda>0$ solutions.

The general procedure is to start with a solution of the
Minkowski equations of motion, and then to continue the coordinates
in such a way so that the signature of space becomes $(+--------)$.
Next, we can send $g_{\mu\nu} \rightarrow -g_{\mu\nu}$ and we are done.

In particular, if we take $\phi=0$, and $T(w) = \bar{T}(\bar{w}) = 0$ we get the inflationary patch.
For $\phi=-2 \log[1+ w \bar{w}]$ we get global de Sitter space,
\be ds^2 = -dt^2 + 4 \cosh^2 t \frac{dwd\bar{w}}{ (1+w\bar{w})^2 } .
\ee
Similarly, when we insert this into the $\Lambda<0$ equation
we get global Euclidean de Sitter in the form
\be ds^2 = dt^2 + 4 \sinh^2 t \frac{dwd\bar{w}}{ (1+w\bar{w})^2 } .
\ee
(Recall that $ds^2 =  \frac{dwd\bar{w}}{ (1+w\bar{w})^2 }$ is the
metric on a two sphere, up to a factor of four.)
In the Euclidean anti-de Sitter case, the metric is defined
for $t\geq 0$ and $t=0$ is a smooth point.

If we take $\phi=-2 \log[(w-\bar{w})/i]$, we find in the de Sitter
case
\be ds^2 = -dt^2 + 4 \sinh^2 t \frac{dwd\bar{w}}{ ( 2 {\rm Im}(w))^2 } ,
\ee
and in the AdS case
\be ds^2 = dt^2 + 4 \cosh^2 t \frac{dwd\bar{w}}{ ( 2 {\rm Im}(w))^2 } .
\ee
This is the usual metric on the upper half plane that is
$SL(2,Z)$ invariant, so once we divide by arbitrary finite
subgroups of $PSL(2,Z)$ the same metric describes locally
the dS (AdS) solution where the Riemann surface has genus
$g>1$. The AdS solution is well-known, and applies to any
dimension, where the $g>1$ Riemann surface is replaced by
an Einstein metric with negative curvature (see e.g. \cite{wittenyau}).

The de Sitter solution has a cosmological contraction/expansion
singularity. It is not clear whether the AdS spaces with disconnected boundary can be made sense of quantum mechanically.   Witten and Yau claim \cite{wittenyau}
that the AdS theory does not make sense usually if the space on which
the dual CFT lives has negative curvature (as it would in these cases).     In the AdS$_5$/CFT$_4$ duality  this is because the conformal coupling of the Ricci scalar to the scalars of the dual theory would cause the action to be unbounded from below. On the other hand, it makes perfect sense to
consider a $1+1$ dimensional CFT on a Riemann surface of arbitrary
genus, and understanding its holographic dual might give important
clues about a dS/CFT duality.

The $w,\bar{w}$ part of the metrics can be written in matrix
notation
as
\be
e^{2t+\phi} \left( \begin{array}{cc} 0 & 1/2 \\ 1/2 & 0 \end{array}
\right) \left[ \left( \begin{array}{cc} 1 & 0 \\ 0 & 1 \end{array}
\right) \pm \frac{1}{2} e^{-2t-\phi}
\left( \begin{array}{cc} -R & -(\bar{T}^{\phi}-\bar{T}) \\
 -(T^{\phi} - T) & - R \end{array} \right) \right]^2
\ee
with the plus sign being appropriate for de Sitter.
For de Sitter, the metric degenerates at
\be \label{deg1}
2e^{2t} = e^{-\phi}(R \pm |T^{\phi}-T|),
\ee
and for AdS at
\be \label{deg2}
2e^{2t} = e^{-\phi}(-R \pm |T^{\phi}-T|) .
\ee
The previous examples had $T^{\phi}=T=0$. In general,
we see that when the AdS solution is smooth the dS
solution degenerates and the other way around.

Now we
compute the on-shell action for these classical solutions. The
Einstein action plus the extrinsic curvature term
for the dS solutions $-dt^2 + h_{ij} dx^I dx^j$ is
evaluated between $t=t_0$ and $t=t_1$ as
\be
S=
\frac{i}{16 \pi G} \left\{
 \int_{t_0}^{t_1} (-4\sqrt{h})
 + [2 \partial_t \sqrt{h} ]_{t_0}^{t_1} ] \right\}
\ee
and for $AdS$ case we obtain the same answer except that $i$ is
replaced by $-1$.
For dS, we also get that
\be \sqrt{h} = \frac{1}{8} e^{2t+\phi} ( (2- e^{-2t- \phi} R)^2 -
e^{-2 t - \phi} |T^{\phi}-T|^2)
\ee
and for AdS we need to send $R\rightarrow -R$. (Here the measure $dwd\bar{w}$ is taken
to be $d{\rm Re}(w) d{\rm Im}(w)$. )
We obtain
\bea
S & = & \frac{i}{32\pi G} \int dw d\bar{w} \left[
2 e^{2t_1+\phi} +4 R t_1 +\frac{1}{2} e^{-2t_1-\phi}
 (|T^{\phi}-T|^2 -R^2) \right. \nonumber \\
 &  & \left. \qquad -
2 e^{2t_0+\phi} -4 R t_0 -\frac{1}{2} e^{-2t_0-\phi}
 (|T^{\phi}-T|^2 -R^2)\right]  .
\label{onshell}
\eea
There are two divergences as $t_1\rightarrow \infty$. These can
be canceled by local counterterms, because
one can easily show that $\sqrt{h} h^{ab} R(h)_{ab} = 4 R = 4\partial
\bar{\partial} \phi$.

We see that by adding boundary terms at both boundaries at $t=t_1$
and $t=t_0$ (which we imagine going to $\pm \infty$), the action
becomes completely zero. (The same result was obtained by \cite{mann} following
the procedure of \cite{brownyork,stressvp,bdbmds,klemm1,pfr}.)

We would have naively expected the Liouville action to appear. That this
does not happen may seem puzzling, because the Liouville action is
needed to reproduce the conformal anomaly, and the arguments for
the conformal anomaly given e.g. in \cite{hensken}
appear to be valid for de Sitter as well. However, a more precise
inspection shows that this is not exactly true. The conformal anomaly arises
from an on-shell action like (\ref{onshell}) from changing $\phi \rightarrow
\phi + \epsilon$ and $t\rightarrow t-\epsilon/2$. Under this change of
coordinate, we pick up a contribution $\int 4 \epsilon R$ from $I^+$,
but at the same time a contribution $-\int 4 \epsilon R$ from $I^-$.
These two contributions cancel each other, and therefore no conformal
anomaly is seen at this level. In the usual AdS case, the conformal anomaly
arises because the second boundary at $I^-$ is absent, and one
assumes that the metric is smooth in the interior and yields
no further divergent contributions. In the degenerating cases with $\Lambda <0$ that were described above, the precise region of integration
for the coordinate is quite complicated: the integration
over $t$ runs from $+\infty$ to the point where the metric degenerates,
which is given by (\ref{deg2}). The Liouville action should arise in
that case due
to this peculiar region of integration of $t$, but we have not
checked this; see \cite{ss} for a related calculation.

In any case, for de Sitter we find that the on-shell action is zero,
since the conformal anomalies on $I^-$ and $I^+$ cancel each other.
Therefore, the right interpretation of this zero should be that it
represents the difference of the Polyakov action for two-dimensional
gravity on the two boundaries,
\be \label{qa20}
 S \sim \Gamma_{I^+}(\hat{g}^+)
- \Gamma_{I^-}(\hat{g}^-)
\ee
For solutions of the equation of motion $S$ vanishes, just as it did
in the case of free scalar fields in the S-matrix picture once
we identified the in and out operators. However, once we
view the two boundaries as independent, (\ref{qa20}) represents the
two Polyakov actions on each boundary, that only cancel each other
once we identify the boundaries and take a classical solution of
the equations of motion. Thus (\ref{qa20}) seems to be the natural
effective action from the S-matrix point of view, and it would
be interesting to verify it directly using an S-matrix type calculation.

\subsubsection{Further properties of the general solution}

By looking at (\ref{deg1}), we can analyze when and whether the
metric degenerates at some point in the interior. This does not
necessarily mean that the space is singular, it is also possible
that we need another set of coordinates to describe the global
structure of the solution. Nevertheless, it is interesting to see
when this happens. If we take the boundary of $dS_3$ to have the
topology of a two-sphere, than the scalar curvature is negative,
and there is no holomorphic quadratic differential. Therefore, we
see from (\ref{deg1}) that the metric degenerates only if for some
point on $S^2$
\be R\pm|T^{\phi}|\geq 0 .\label{cond1}
\ee
For a pure two-sphere which corresponds to global de Sitter space,
we took $\phi=-2\log(1+w\bar{w})$, and it follows that
$R=-2/(1+w\bar{w})^2$ and $T^{\phi}=0$. Therefore, the left hand
side of (\ref{cond1}) is always negative and the metric does not
degenerate, as expected. A small deformation of the metric on the
two-sphere is still allowed, as long as the corrections decay
sufficiently fast as $w,\bar{w}\rightarrow \infty$ so that they do
not dominate over $R=-2/(1+w\bar{w})^2$ for large $w,\bar{w}$. As
the perturbations become larger, they reach a point where the
metric degenerates somewhere in the bulk.

It is also interesting to compute the Brown-York stress tensor for
the general solution. Using the prescription of
\cite{bdbmds} we find that the Brown-York stress-tensor at
$I^+$ is given by
\be \label{by}
T^{BY}_{ij} = -\frac{1}{8\pi}
\left( \begin{array}{cc} -2T^{\phi} & \frac{1}{2} R  \\
 \frac{1}{2} R & - 2 \bar{T}^{\phi} \end{array} \right) .
\ee
We see that the Brown-York stress tensor contains the Liouville
stress tensor $T^{\phi}$. Although we normally use the Brown-York
stress tensor to define conserved quantities, if the boundary is a
generic deformation of the two-sphere there are no isometries of
the boundary geometry and there are no obvious conserved
quantities associated to this Brown-York tensor. Nevertheless, it
seems to know about the masses of e.g. de Sitter like solutions
with point like defects. These can be constructed by taking
\be \phi=\log\left( \frac{\gamma^2 w^{\gamma-1}
\bar{w}^{\gamma-1}}{(1+w^{\gamma} \bar{w}^{\gamma})^2} \right)
\ee
which corresponds to taking a two-sphere from which a piece has
been cut out with the boundaries identified, so that conical
singularities appear at the north and south poles. This Liouville
field has
\be R = - 2
\frac{\gamma^2 w^{\gamma-1} \bar{w}^{\gamma-1}}{(1+w^{\gamma}
\bar{w}^{\gamma})^2} + \pi (\gamma-1) \delta^2(w,\bar{w})
\ee
and
\be T=\frac{-1+\gamma^2}{2w^2}. \ee
The curvature $R$ indeed has a singularity at $w=\bar{w}=0$, and
the $L_0$ component of $T$ is closely related to the mass of the
corresponding conical defect spaces as computed in
\cite{bdbmds}.

It would be interesting to generalize the above observations,
study their connection with the mass conjecture of
\cite{bdbmds}, and the relation with the results of
\cite{klemm}.

\subsubsection{On-shell action and entropy}

We now discuss the connection between the
on-shell action and the entropy of dS space; according to
\cite{banksfischler}, the entropy is related the dimension of the
Hilbert space of the dual theory, although this is not the point
of view that we will take.
In the Euclidean case, the identification of the entropy with
the on-shell value of the action
seems to be natural from the Gibbons-Hawking Euclidean approach \cite{gh1}.
Note that the euclidean on-shell action as computed above
yields precisely the Bekenstein-Hawking entropy \cite{bekhaw}.
If we take time imaginary with period $2\pi l$
\cite{gh1, nappi}, and insert this into
(\ref{onshell}), we see that the the exponentials of time cancel
each other, and the action becomes $-(l/4G) \int d^2 w R$. With the
conventions used here, the integral of $R$ is $-2\pi$, and the action
becomes $S=\pi l/2G$.

\subsection{The Chern-Simons perspective}

In this section we revisit the case of pure gravity in $2+1$
dimensions from the Chern-Simons perspective. In some
sense, the duality between the $2+1$-dimensional Chern-Simons
theory and the $1+1$ WZW model is a prototype example of
an AdS/CFT correspondence, and understanding the Chern-Simons
theory for de Sitter space may lead to a better understanding
of the nature of the dS/CFT correspondence.

In particular, we will study the nature of the
boundary conditions that we have to impose on the Chern-Simons
gauge fields. The relevant Chern-Simons theory has gauge group
$SL_2(C)$. It turns out that the boundary conditions depend on
the coordinate system and corresponding patch of de Sitter space
we are considering.
In the inflationary patch, the situation
is very similar to the analogous situation discussed
in the AdS/CFT correspondence \cite{adsduality2, lorentzian,lorentzian2}.
The global patch is different, due to the two boundaries we
have to consider. The final conclusion we will reach is similar
to the conclusion reached in the previous section; for solutions
of the equation of motion, the on-shell action is essentially trivial.
Again, the difference with the AdS case lies in the assumption
that the solution is regular in the interior. This is why the
derivation of the Liouville action from Chern-Simons theory recently
given in \cite{banados} does not apply to de Sitter space.
We will also briefly discuss the gauge fields relevant
for Kerr-de Sitter space.

The $SL_2(C)$ CS theory has been studied by
Witten in \cite{wittencs}.   (Other interesting recent discussions that uses the Chern-Simons approach to explore de Sitter entropy include~\cite{jens, Gov}.)   It has level $t=k+is$, with $k$ integer (and level $\bar{t}=k-is$ for the complex conjugate gauge field.)
The usual $\Lambda>0$ $2+1$ Einstein action
arises when $k=0$ and $s$ is real. The action is real
if $s$ is real. The theory is also unitary for
purely imaginary $s$, in which case there is a different
inner product and the theory is related to Euclidean
gravity with $\Lambda<0$.   We will come back to this observation
in section 3. Explicitly,
the action of gravity in $2+1$ dimensions with $\Lambda >0$ is
given by
\be \label{csact} S = \frac{is}{4\pi} (I_{CS}(A) - I_{CS}(\bar{A}) ),
\ee with \be I_{CS}(A) = \int d^3 x {\rm Tr} (A\wedge dA +
\frac{2}{3} A^3) \ee the usual Chern-Simons functional. The gauge
field $A$ takes values in $SL(2,C)$, and $\bar{A}$ is the complex
conjugate of $A$. We have not yet included boundary contributions,
and the first step will be to determine a reasonable set of
boundary conditions. The vielbein and spin-connections are given
by \be e=(A-\bar{A})/(2i) , \qquad \omega=(A+\bar{A})/2 . \ee

The following discussion is similar to the discussion for AdS space
given in \cite{banados}, but some of the details will be different.
A first important remark is that for solutions of the equations of
motion, the time-dependence of the gauge field
can be encoded in a gauge transformation with \be
U = \left( \begin{array}{cc} \cosh(t/2) & -i \sinh(t/2)  \\
i \sinh(t/2) & \cosh(t/2) \end{array} \right) \label{defu}
\ee
so that
\be \label{conn}
A = U^{-1} \left( \begin{array}{cc}
 \alpha^3/2 & \alpha^+ \\ \alpha^- & - \alpha^3/2 \end{array}
\right) U + U^{-1} \partial_t U \ee with $\alpha^{3},\alpha^{\pm}$
time independent. Actually, it is convenient to work with a
different basis than the standard one. Introduce \be N^{\pm} =
\frac{1}{2} (\alpha^+ + \alpha^- \mp i \alpha^3), \qquad N^3 =
\frac{1}{2}(\alpha^+ - \alpha^-), \ee and the following basis for
$SL(2,C)$,
\be T^+ = \left( \begin{array}{cc} -i/2 & 1/2 \\ 1/2 &
i/2 \end{array} \right),\quad T^- = \left( \begin{array}{cc} i/2 &
1/2 \\ 1/2 & -i/2 \end{array} \right),\quad T^3 = \left(
\begin{array}{cc} 0 & -i/2 \\ i/2 & 0 \end{array} \right) .
\ee
This basis obeys the usual commutation relations, $[T^+,T^-]=2T^3$
and $[T^3,T^{\pm}]=\pm T^{\pm}$. The connection (\ref{conn})
becomes
\be A = (dt + 2 i N^3) T^3 + e^t N^+ T^- + e^{-t} N^- T^+ =
U^{-1} \tilde{A} U + U^{-1} \partial_t U
\ee
with
\be
\tilde{A} = 2 i N^3 T^3 + N^+ T^- +  N^- T^+ .
\ee
The vielbein on spatial slices as $t\rightarrow \infty$ is
therefore \be e^{+\infty}  = \lim_{t\rightarrow + \infty} e^{-t}
\frac{A-\bar{A}}{2i} =\frac{1}{2i} ( N^+ T^- - \bar{N}^+ T^+), \ee
and similarly \be e^{-\infty}  = \lim_{t\rightarrow - \infty}
e^{t} \frac{A-\bar{A}}{2i} =\frac{1}{2i} ( N^- T^+ - \bar{N}^-
T^-). \ee The metrics at $t\rightarrow \pm \infty$ are obtained
from ${\rm Tr}(ee)$ and we find $\frac{1}{2} N^+ \bar{N}^+$
respectively $\frac{1}{2} N^- \bar{N}^-$.

\subsubsection{Global patch}

To obtain the most general three dimensional solution we discussed in
section~2.3  we  take
\be \label{gensol1}
N^+ = e^{\phi/2} dw, \qquad
N^- = \frac{1}{2} e^{-\phi/2} (( T^{\phi}-T(w))dw + R d\bar{w} ) .
\ee
In addition, $A$ needs to be flat, so that equivalently
\be
2 i N^3 T^3 + N^+ T^- + N^- T^+
\ee
needs to be a flat connection. This implies that $A$ is pure gauge
if $\pi_1(M)=0$. If $\pi_1(M)\neq 0$ we need to include holonomies around
the non-contractible cycles in space-time. This may be relevant
if we want to do the black hole and conical defect cases, where the topology is
$S^1 \times R^2 $ instead of $S^2 \times R$.

Flatness implies
\bea
dN^3 - i N^- \wedge N^+ & = & 0 \nonumber \\
dN^+ - 2 i N^3 \wedge N^+ & = & 0 \nonumber \\
dN^- + 2 i N^3 \wedge N^- & = & 0 .
\eea
These are solved by
\be \label{gensol2}
N^3 = \frac{1}{4i} (\bar{\partial} \phi d\bar{w} -
 \partial \phi d w ) .
\ee

In particular, for global de Sitter space we had $\phi=-2 \log(1+w
\bar{w})$, and the flat gauge field $\tilde{A}$ reads
\be
\tilde{A} = \left( \begin{array}{cc} \frac{i}{2} & \frac{1 -
i\bar{w}}{2} \\ \frac{1+i \bar{w}}{2} & -\frac{i}{2}
\end{array} \right) \frac{dw}{1+w\bar{w}} +
\left( \begin{array}{cc} \frac{i}{2} & \frac{-1 + i{w}}{2} \\
\frac{-1-i {w}}{2} & -\frac{i}{2}
\end{array} \right) \frac{d\bar{w}}{1+w\bar{w}} .
\ee

These gauge fields determine the behavior of the connection $A$ as
$t\rightarrow \pm \infty$, and therefore are intimately related to
the boundary conditions we want to impose on $A$. The form of
$A_w$ does not appear immediately related to the form we need to
perform a Hamiltonian reduction from an $SL_2(C)$ WZW model to a
Liouville-like theory. The Sugawara stress tensor of such a WZW
model living on the boundary would be proportional to $\tr (A_w
A_w)$, which in this case is equal to $\bar{w}^2/2(1+w\bar{w})^2=
\frac{1}{8} \partial\phi \partial \phi$. The nonvanishing of this
stress tensor can be viewed as evidence that de Sitter space in
global coordinates should not be identified with the vacuum state
in some conformal field theory, in agreement with the picture we
develop later in this paper.

\subsubsection{Inflationary patch}

It is straightforward to obtain $\tilde{A}$ and $A$ in the
inflationary patch, using (\ref{gensol1}) and (\ref{gensol2}), as
the inflationary patch corresponds to $\phi=0$. In particular, for
$\tilde{A}$ we find
\be
\tilde{A} = T^- dw = \frac{1}{2} \left( \begin{array}{cc} i & 1 \\
1 & - i \end{array} \right) .
\ee
Since $T^-$ is conjugate in $SL_2(C)$ to the matrix
\be \left(\begin{array}{cc} 0 & 0 \\
1 & 0 \end{array} \right)
\ee
this is of form needed to perform a Hamiltonian reduction from an
$SL_2(C)$ WZW model to a Liouville-like theory, similar to what
happens in de AdS case \cite{vandrieletal}. This can be further
illustrated by looking at the form of $\tilde{A}$ under a Virasoro
transformation. In other words, we transform $\tilde{A}$ with the
Virasoro vector field
\be \label{virvec}
L_{\xi}= -\xi \partial_w + \frac{1}{2} \xi' \partial_t
-\frac{1}{2} e^{-2 t} \xi'' \partial_{\bar{w}} + {\rm c.c.}
\ee
where $\xi(w)$ is holomorphic. Since gauge fields transform as
one-forms, and $\tilde{A}_t=\tilde{A}_{\bar{w}}=0$, $\tilde{A}$
transforms into
\be
\tilde{A} \rightarrow (1-\xi') T^- dw .
\ee
An infinitesimal gauge transformation with parameter
$\epsilon=\xi' T^3 + \xi'' T^+$ changes this, to first order in
$\xi$, into
\be
\tilde{A}_w \rightarrow (1-\xi') T^- + [T^-,\epsilon]+
\partial_w \epsilon = T^- + \frac{1}{2} \xi''' T^+ .
\ee
Since the transform of the stress-tensor contains a term $\xi'''$,
this transformed $\tilde{A}$ is conjugate to a matrix of the form
\be \left(\begin{array}{cc} 0 & cT \\
1 & 0 \end{array} \right)
\ee
with $c$ some constant and $T$ the stress tensor. Indeed, this is
exactly of the form one needs to perform Hamiltonian reduction of
the $SL_2(C)$ WZW model to a theory of induced $1+1$ dimensional
gravity on the boundary.

\subsubsection{Kerr-de Sitter solution}

We take the metric for the Kerr-de Sitter space \cite{deserjackiw,
park} discussed in our previous paper \cite{bdbmds}
\be \label{metric2}
ds^2 = -\frac{(r^2 + r_-^2)(r_+^2 - r^2)}{r^2} dt^2 +
\frac{r^2}{(r^2 + r_-^2)(r_+^2 - r^2)}dr^2 + r^2 (d\phi +\frac{r_+
r_-}{r^2} dt)^2
\ee
Following \cite{park} we define
\be
r^2 = \sinh^2 \rho r_-^2 + \cosh^2 \rho r_+^2
\ee
which yields the metric
\be
ds^2 = -d\rho^2 + \cosh^2 \rho (r_- dt + r_+ d\phi)^2 + \sinh^2
\rho (r_+ dt - r_- d\phi)^2 .
\ee
This is again of the general form given in section~2.3, with
$\rho$ playing the role of time. Thus, we can use a gauge
transformation with $U$ as in (\ref{defu}), with $t$ replaced by
$\rho$, to remove the $\rho$ dependence from the $SL_2(C)$ gauge
field that describes the Kerr-de Sitter metric. The result is
\be \label{eqa}
\tilde{A} = \left( \begin{array}{cc} 0 & \frac{1}{2}
(r_- - i r_+) (d\phi-i dt) \\
\frac{1}{2} (r_- - i r_+) (d\phi-i dt) & 0
\end{array}
\right). \ee This is the equation that is relevant to get the
$L_0$ and $\bar{L}_0$ eigenvalues for the Kerr de Sitter
spacetimes. In the present setup, we again identify $A$ with the
currents of a WZW model, and ${\rm tr}(AA)$ with the Sugawara
stress tensor of that WZW model. Since ${\rm tr}(AA)\sim (r_- - i
r_+)^2$ we indeed recover (up to factors of $i$) the mass and
angular momentum of the Kerr de Sitter black hole.

Therefore, this type of analysis in terms of gauge fields provides
valuable information about a putative dual theory. For the gauge
field story of the BTZ black hole \cite{btz} consult
\cite{banados}.

\subsubsection{On-shell action}

Let us try to compute the value of the on-shell action in the
Chern-Simons setup.   This requires us first of all to determine the
appropriate boundary terms. Following \cite{banados}, we will
impose boundary conditions that preserve the asymptotic form of
the metric. As $t\rightarrow +\infty$, we therefore want to keep
the form of $N^+$ fixed, which implies
 \be \delta {\rm Tr}(T^+ A) =0,\quad t\rightarrow +\infty \ee
and similarly as $t\rightarrow -\infty$, we want to fix $N^-$, so
that \be \delta {\rm Tr}(T^- A) =0,\quad t\rightarrow -\infty .\ee

Also, in order to make sure there is no $dtdw$ or $dtd\bar{w}$
term in the metric, we also will impose
 \be \label{pq} {\rm Tr}( T^3 (A-\bar{A}) ) =0 .\ee

To examine the structure of the action, we decompose
 \be A= A^+ T^- + A^- T^+ + A^3 T^3 \ee
 and therefore
 \be \bar{A} = \bar{A}^+ T^+ + \bar{A}^- T^- - \bar{A}^3 T^3 \ee
 using the explicit form of $T^i$ and the fact that $\bar{A}$ is
 the complex conjugate of $\bar{A}$.

Dropping the factor of $s/4\pi$, the variation of the CS action
yields a boundary term
 \be \delta S =
 -i \int_{\partial M} {\rm Tr} (A^- \delta A^+ + A^+ \delta A^-
  + \frac{1}{2} A^3 \delta A^3) + \mbox{\rm c.c.} \ee

Now as $t\rightarrow \infty$, we want to find a boundary term
proportional to $\delta A^+$ so that its vanishing implies the
vanishing of the relevant  boundary terms. This can be
accomplished by adding
 \be S_1= i \int_{\partial M} {\rm Tr} (A^+ \wedge A^-) +
 \mbox{\rm c.c.} \ee
To take care of (\ref{pq}), we add
 \be S_2= \frac{i}{2} \int_{\partial M} {\rm Tr}
  (A^3 \wedge \bar{A}^3) .
  \ee
The total variation becomes
 \be
 \delta (S+S_1+S_2) = \int_{\partial M}
 {\rm Tr}(-2i A^- \delta A^+ +2 i \bar{A}^- \delta \bar{A}^+
 -\frac{i}{2} (A^3 + \bar{A}^3)\delta (A^3 -\bar{A}^3))
 \ee
 which is indeed consistent with the boundary conditions
 as $t\rightarrow +\infty$.

On the other hand, on the other boundary at $t\rightarrow
-\infty$, we want to put $\delta A^-=0$, and this requires us to
take $S-S_1+S_2$ instead of $S+S_1+S_2$. Notice that in all this
discussion $\partial M$ carries the standard orientation induced
from that of $M$ and the outward normal at each boundary.

The bulk action, for a solution of the equation of motion, equals
 \be S= i \int_{M} {\rm Tr}( A^3 A^+ A^- - \bar{A}^3 \bar{A}^+
  \bar{A}^-)
  \ee
where we used flatness of $A$ to write the CS action as
$-\frac{1}{3} \int A^3$. The components of $A$ are given by
 \bea
 A^3 & = & dt + 2 i N^3 \nonumber \\
 A^+ & = & e^t N^+ \nonumber \\
 A^- & = & e^{-t} N^- .
 \eea

The total value of the on-shell action is
\bea
S & = & i \int_M {\rm Tr}( A^3 A^+ A^- - \bar{A}^3 \bar{A}^+
  \bar{A}^-)
\nonumber \\
& & \pm i \int_{\partial M} {\rm Tr} (
A^+ \wedge A^- - \bar{A}^+ \wedge \bar{A}^-) \nonumber \\
& & + \frac{i}{2} \int_{\partial M} {\rm Tr} (
A^3 \wedge \bar{A}^3 ) .
\label{totalcs}
\eea

This seems to give (almost) zero. Notice that $A^+ \wedge
A^-$ is proportional to the curvature, and the middle term gives
something proportional to the curvature. The bulk term can be
integrated and gives a logarithmically divergent term proportional
to the curvature R. This is subtracted upon regularization, or
replaced by a constant times the curvature. The last boundary term
vanishes because it is propotional to $N^3 \wedge N^3$ which
vanishes (the field $\phi$ is real).   (Note that the results here are a little different from those obtained earlier using the Einstein action for gravity since we are treating the boundary terms differently and are not being careful about finite terms left over from removing the logarithmic bulk diveregence.)

The the situation is completely different from AdS, where a similar calculation gives the Liouville action. This is among other things due to the fact that the boundary terms here are
different from the ones in AdS \cite{banados}. Still, taking
boundary terms similar to the ones in \cite{banados} does not
improve the situation very much. The main difference is that de
Sitter space has two boundaries, whereas AdS has a single
boundary. The Liouville action arises in \cite{banados}
precisely because one makes an assumption about the regularity of
the solution in the interior, so that the only boundary
contributions in partial integration is from the AdS boundary,
although the coordinates one uses are not only bounded by the
boundary of AdS but also by the region in the interior where the
metric in those coordinates degenerates, as given by (\ref{deg1}).
It may be possible to similarly recover the Liouville action on
the inflationary patch of de Sitter space using an assumption
about the regularity at the horizon of the inflationary patch, but
we have not verified this; see e.g. \cite{klemm} for a
construction of this type.

The situation is similar to what we discussed in
section~2.3. The Liouville action does not arise from a naive
calculation of the on-shell action. However, treating the the two boundaries as separate surfaces and carefully removing the bulk divergence in (\ref{totalcs}) by adding suitable boundary terms we would expect to find that the total action vanishes, but consists of two separate Liouville pieces on different boundaries that cancel.

Altogether the main lesson from this section is that it is rather
subtle to impose the right boundary conditions, but for any choice
of boundary condition the naive analysis of the bulk action as a function of boundary data yields an uninteresting result.  However, by examining the contributions from the the two boundaries of de Sitter space separately we can get more interesting answers that agree with S-matrix type
calculations in the bulk. This structure is
evidence that if there is a  dual theory it should be two (possibly entangled) CFT's living on
separate spaces rather than one. In the next section we will argue
that these CFT's are of a novel form, with different hermiticity
conditions from what we are used to in conformal field theory. In
section~4 we will then combine all these observations to arrive at
a proposal of what a dual description of de Sitter space could
look like.

\section{Hermiticity and the inner product}

If there is a CFT dual to de Sitter space there are also indications that it will have unconventional hermiticity and unitarity properties.    For example, conventional unitarity suggests a upper bound on particle masses in de Sitter space \cite{andyds} in an analog of the Breitenlohner-Freedman bound \cite{breitfreed} in AdS.   Such a bound is difficult to understand from a physical point of view.   However, there is reason to believe that a unitary theory with SL(2,C) symmetry is relevant for three dimensional de Sitter space.

First of all, in the Chern-Simons formulation of 3d gravity, the $SL(2,C)$
CS theory \cite{wittencs} with level $t=k+is$, with $k$ integer (and level $\bar{t}=k-is$ for the complex conjugate gauge field) describes both de Sitter space and Euclidean AdS. The usual $\Lambda>0$ $2+1$ Einstein action arises when $k=0$ and $s$ is real. The action is real if $s$ is real. The theory is also found to be unitary for purely imaginary $s$, in which case there is a different
inner product and the theory is related to Euclidean gravity with $\Lambda<0$.   Therefore, since Euclidean AdS is related to a Euclidean CFT with conventional inner product and hermiticity conditions, we expect some unusual features in the inner product for a dual to de Sitter.

A guess for the unusual inner product follows from the relation
between CS theory in the bulk and WZW theory on the boundary. The
CS action relevant for $\Lambda>0$ is given in (\ref{csact}). On a
manifold with boundary, the CS action is related to a WZW theory
living on the boundary \cite{wittenjones,elitzuretal}. The CS
action for a compact gauge group reduces to a chiral WZW theory on
the boundary. With gauge group $SL(2,C)$, one would therefore
expect a chiral $SL(2,C)$ WZW theory on the boundary. However,
there are two terms in the action (\ref{csact}). The first term
gives rise to a chiral $SL(2,C)$ WZW theory, the second one to an
antichiral WZW theory (\cite{wittencs}, see also
\cite{klemm}). However, the gauge field $\bar{A}$ was the complex
conjugate of $A$, and therefore the antichiral WZW theory is
related by complex conjugation to the chiral WZW theory. This is
quite distinct from the case with $\Lambda<0$, where $A$ and
$\bar{A}$ are independent $SL(2,R)$ gauge fields. On the boundary
we obtain independent chiral and antichiral $SL(2,R)$ WZW theories
that combine into one standard $SL(2,R)$ theory.

In the $\Lambda>0$ case that is relevant for de Sitter space, we
seem to find a WZW-like theory on the boundary, but one with
different hermiticity conditions. Namely, the hermiticity
conditions also involve an exchange of left and right movers.  (See, for example, \cite{jens}.)  One
may check that on the level of zero modes, such hermiticity
conditions are compatible with unitary representations of
$SL(2,C)$.  In fact, this is in perfect agreement with what we find for a
free scalar field in de Sitter space; as we will elaborate in the
next section, the Hilbert space of a free massive scalar field in
de Sitter space consists of unitary representations of $SL(2,C)$.

An $SL(2,C)$ WZW theory was also used in \cite{juanandy1} in an
attempt to explain the entropy for de Sitter space a la Carlip
\cite{carlip}. That $SL(2,C)$ theory did however not live on the
boundary of de Sitter space and is not obviously related to our
discussion here.

A further motivation for these novel hermiticity conditions
follows from computations of $L_0$ and $\bar{L}_0$ for various
solutions of the Einstein equations with $\Lambda>0$.
Given the asymptotic 2d euclidean conformal symmetries of the
$\ds{3}$, the boundary stress tensor formalism can be used to
compute the eigenvalues of $L_0$ and $\bar{L}_0$ for a variety of
conical defect and spinning defect (Kerr-dS)
spacetimes~\cite{bdbmds}.  These eigenvalues are not consistent
with a unitary representation of SL(2,R) $\times$ SL(2,R) as we
might have expected.   Rather, there are consistent with the
(unitary)  principal series representations of $SL(2,C)$.

So far, attempts to find de Sitter solutions in string theory
frequently involve unusual reality condition as for example in the
work of Hull \cite{hull}. There, de Sitter always seems to arise
in cases where reality conditions are unusual and wrong sign
fields appear\footnote{Another problem with this set-up is the
fact that it is hard to talk about de Sitter gravity in lower
dimensions; solutions found in \cite{hull} always involve de
Sitter space times a non-compact manifold.}. These wrong signs may
be natural from the point of view of a unitary theory with $SL(2,C)$
symmetry. As we will speculate in the conclusions,  this might 
all makes sense in a new class of string theories, where the
hermiticity conditions are not the usual ones.

With all of these points in mind we will now explore a simple toy model
of the unusual Hermiticity conditions for 2d CFT suggested
by the Chern-Simons analysis of de Sitter space described above.

\subsection{A toy model}
\label{sec:toymodel}
We arrive at our toy model by considering the simplest Chern-Simons theory, that of $U(1)$.
On the boundary of a three-manifold, it gives rise to a chiral
boson. However, in case of $dS_3$, we have a complex $SL(2,C)$
gauge field and its complex conjugate, and a certain reality
condition relating the two.   On the boundary, this becomes two
chiral WZW theories, each with $SL(2,C)$ current algebra. The
Euclidean AdS reality condition tells us that both chiral currents
should be viewed as $SL(2,R)$ currents, rather than $SL(2,C)$
currents, and subsequently the two chiral $SL(2,R)$ models can be
combined in a single $SL(2,R)$ WZW model. The other reality
condition, relevant for $dS_3$, is a suitable relation between the
two currents, roughly of the form
\be
J_L^{\dagger}=J_R
\ee
where the subscripts $L,R$ denote the left and right moving sector.  (See, for example, \cite{jens}.) 

In our toy model, we consider $U(1)$ currents with the reality condition
\be\label{e1}
J_L^{\dagger} = i \, J_R
\ee
which leads to the Hermiticity conditions on the Virasoro
generators that we expect to be appropriate for the de Sitter
case. Thus, we consider a complex left-moving chiral boson $\phi$
and a complex right moving chiral boson $\chi$. The currents are
\be
J_L = i \partial \phi = \sum \alpha_{n} z^{-n-1},
\qquad
J_R = i \bar{\partial} \chi = \sum \beta_{n} {\bar{z}}^{-n-1}
\ee
and we will assume the usual commutation relations between
$\alpha_n$ and $\beta_n$,
\be \label{e2}
[\alpha_m,\alpha_n]=[\beta_m,\beta_n] = m \delta_{m+n,0},
\ee
but not the usual hermiticity conditions. The hermiticity conditions
that follow from (\ref{e1}) suggest to take
\be \label{e3}
\alpha_n^{\dagger} = i\beta_n, \qquad
\beta_n^{\dagger} = i\alpha_n .
\ee
A different but equivalent choice of hermiticity conditions is to
take $\alpha_n^{\dagger}=\beta_{-n}$, and $\beta_n^{\dagger}=
\alpha_{-n}$. This shows that the new hermiticity conditions are
obtained by combining the usual ones with an exchange of left and
right movers.

The Virasoro generators should be a normal-ordered version of the
usual answer $T=-\frac{1}{2} \partial \phi \partial \phi$. The
normal ordering depends on a choice of vacuum etc. We will not
discuss this issue yet, but simply note that with a suitable
normal ordering the hermiticity conditions become\footnote{An
equivalent set of hermiticity conditions would be to take
$L_k^{\dagger}=\bar{L}_{-k}$ instead.}
\be \label{e4}
L_k^{\dagger} = -\bar{L}_{k}
\ee
which then also implies that the central charges should obey
\be
\label{e5}
c_L^{\ast} = -c_R
\ee
which is compatible with a purely imaginary central charge for
both the left and right movers. (This type of relation was
encountered in our discussion of the mass of de Sitter space in
\cite{bdbmds}.) In the following we will refer to a Virasoro
algebra with these Hermiticity conditions as a ``Euclidean
Virasoro algebra" since it arose from considering possible
Euclidean CFTs dual to de Sitter space.

The hermiticity conditions (\ref{e4}) give rise to unitary
representations of $SL(2,C)$. In general, given some Lie algebra
and some anti-linear operator $\dagger$ that squares to one, the
algebra that will be unitarily implemented is the $-1$ eigenspace
of $\dagger$.  One then readily verifies that the combinations
$L_m+\bar{L}_m$ and $i(L_m-\bar{L}_{m})$ with eigenvalue $-1$
under the hermiticity conditions (\ref{e4}) form the $SL(2,C)$ Lie
algebra, and therefore they are unitarily implemented.\footnote{It
is worth remembering why the usual hermiticity condition
$L_m^{\dagger} = L_{-m}$ gives $SL(2,R)$ rather than $SU(2)$ as
unitarily represented subalgebra. Clearly, $L_k^{\dagger} = -L_k$
gives $SL(2,R)$, but the other one seems like the hermiticity
condition for $SU(2)$. This point is a bit subtle, but the
solution is that in terms of Pauli matrices, $i\sigma_1,
i\sigma_2,i\sigma_3$ generate $SU(2)$, whereas
$\sigma_1,\sigma_2,i\sigma_3$ generate $SL(2,R)$, in a somewhat
nonstandard basis. Given $L_m^{\dagger}= L_{-m}$, the $(-1)$
subalgebra is generated by $i\sigma_3 \sim iL_0, \sigma_1\sim
L_1-L_{-1}, \sigma_2 \sim i(L_1+L_{-1})$ and therefore gives a
unitary representation of $SL(2,R)$ rather than $SU(2)$.}

\subsubsection{Representations}

\paragraph{Stone-von Neumann representations: } To study the representations of the oscillator
algebra (\ref{e3}) we construct the hermitian combinations
\be
\delta_n =
\alpha_n + i\beta_n ,\qquad \epsilon_n = i(\alpha_n - i\beta_n).
\ee
The commutation relations between these objects read
\be
[\delta_m,\delta_n]=[\epsilon_m,\epsilon_n]=0,\qquad
[\delta_m,\epsilon_n] = 2 i m \delta_{m+n,0} .
\ee
This is exactly
the same as the usual relations between coordinates and momenta,
and the representations are known from the Stone-von-Neumann
theorem. The unique irreducible unitary representation of the
Hilbert space is
\be \label{e8}
\otimes_{n> 0}
L^2(\delta_n)\otimes|p\rangle  .
\ee
where $L^2(\delta_n,\epsilon_n)$ are the $L^2$ normalizable functions of $\delta_n$ and $\epsilon_n$.
Here, $|p\rangle$ is a representation of the zero modes that satisfies
\be
\label{e7} (\alpha_0 + i\beta_0)|p\rangle = p_1,\qquad i (\alpha_0
- i\beta_0) |p\rangle = p_2
\ee
for real $p_1,p_2$. Thus,
$\alpha_0 |p\rangle = (p_1-ip_2)/2 |p\rangle$ and $\beta_0
|p\rangle = (p_2-ip_1)/2 |p\rangle$.

Actually, it is not quite clear that the Stone-von Neumann
representation is also the unique one in case we have an infinite
number of degrees of freedom, but we have not found any other
obvious representations of the commutations relations.

\paragraph{Highest weight representations: } We could also try to build more conventional representations by imposing the following conditions
\be
\alpha_n|p\rangle = \beta_{-n}|p\rangle, \qquad n>0
\ee
on a highest weight state $|p\rangle$ that obeys
(\ref{e7}). Though interesting, it is not clear
that these states can be organized in a unitary representation.
To define an inner product we need a bilinear form
\be
B( \prod_{n_i}\beta_{n_i}\prod_{m_j}\alpha_{-m_j}|p\rangle,
 \prod_{n_k}\beta_{n_k}\prod_{m_l}\alpha_{-m_l}|p\rangle ) .
\ee
There are several possibilities. First, one can try to define an
inner product compatible with the hermiticity conditions. But then
$B(\alpha_{-1} |p\rangle ,\alpha_{-1} |p\rangle ) = B(|p\rangle,
i\beta_{-1} \alpha_{-1} |p\rangle ) =0$ and the inner product is
not positive definite. One can also define a more conventional
inner product on the highest weight modules, as we do for the
usual free field. This yields a positive definite inner product,
but now the oscillators do not obey the required hermiticity
conditions.  Later we will comment on the relation between these
representations and the Hilbert space of a free scalar field in de
Sitter space.

\subsubsection{Calculations in the toy model}

Calculations in this toy model are not completely straightforward.
One first needs to construct a state in the Hilbert space
(\ref{e8}). The state $|p\rangle$ is by itself not an element of
the Hilbert space, because $1$ is not a square integrable
function. One can of course write down an infinite set of square
integrable functions of the variables $\delta_n$, and use the
corresponding state to do calculations. In particular, one can
compute correlation functions of the field $\delta \phi$ in such a
state. A particularly convenient basis for $L^2(\delta_n)$ is to
use harmonic oscillator wave functions. These are of the form
$(\delta_n + i \epsilon_{-n})^k |\psi_n\rangle$, where
$\psi_n\rangle$ is a state annihilated by the annihilation
operator $\delta_n - i \epsilon_{-n}$. These states are not
invariant under $SL(2,C)$, so we cannot use symmetry principles to
determine the form of two and three point functions as in usual
conformal field theory. It would be interesting to study the
nature of these correlations functions in more detail.

\subsubsection{Virasoro representations}

We return to the representation (\ref{e8}) which has a positive
definite norm and respects the hermiticity conditions.   It is
interesting to to decompose this into representations of the
Virasoro algebra, with the new hermiticity conditions
$L_k^{\dagger} = -\bar{L}_k$.    The representation theory of this
Virasoro algebra should be a suitable generalization of the
representation theory of $SL(2,C)$.

It is convenient to rewrite the Hilbert space slightly using
complex coordinates $z_n= (\delta_n - i \epsilon_n)/\sqrt{n}$.
(Then $z_n$ and $\bar{z}_n$ are rescaled versions of $\alpha_n$
and $\beta_n$.) The Hilbert space is the set of square integrable
functions of $z_n,\bar{z}_n$, times the highest weight state
$|p\rangle$.  For the Virasoro generators we find $L_m\sim \sum_n
z_n \partial/\partial z_{m-n}$ and similarly for their complex
conjugates. This is reminiscent of the form of the Virasoro
generators in the old matrix model \cite{oldmatrix}; perhaps there
is an interesting connection here.

So what are the representations of the Euclidean Virasoro algebra
with hermiticity conditions (\ref{e4})? By analogy with the usual
case a reasonable guess could be that for generic eigenvalues of
the zeromodes, the oscillator Hilbert space furnishes an
irreducible unitary representation of the Euclidean Virasoro
algebra. This is also reasonable because $SL(2,C)$ has its unitary
representations on square integrable functions of a single complex
variable (with standard measure for the principal
representations). If the modes $\{L_{\pm 1},L_0, \bar{L}_{\pm
1},\bar{L}_0\}$ have such a representation, it is quite reasonable
that adding more modes of the Virasoro algebra will require extra
sets of square integrable functions of a complex variable. It
would be interesting to develop this representation theory in more
detail.

In terms of the complex functions, and with a suitable normal
ordering that we still have not specified, the lowest order
Virasoro generators will become\footnote{We employ
$\alpha_n=z_n \sqrt{n}$, $\alpha_{-n}=-\sqrt{n} \, \partial/\partial z_n$,
$\beta_n=-i\sqrt{n} \bar{z}_n$, $\beta_{-n}=-i\sqrt{n} \, \partial/\partial \bar{z}_n$ for
$n>0$.}

\bea
L_{-1} & = & -\sum_{n>0}
\sqrt{n(n+1)} z_n \frac{\partial}{\partial z_{n+1}} -\tilde\alpha_0
\frac{\partial}{\partial z_1} \nonumber \\
L_0 & = & \frac{1}{2} \tilde\alpha_0^2 - \sum_{n>0} n z_n
\frac{\partial}{\partial z_n} \nonumber \\
L_1 & = & -\sum_{n>0} \sqrt{n(n+1)} z_{n+1}
\frac{\partial}{\partial z_{n}} + \tilde\alpha_0 z_1
\eea
where $\tilde{\alpha}_0=(p_1-ip_2)/2$ and we have left out a
normal ordering constant in $L_0$. Similar expressions can be
written down for the antiholomorphic generators.   To be
compatible with the Hermiticity conditions, the normal ordering
constant is chosen as
\be
L_0  =  \frac{1}{2} \alpha_0^2 - \sum_{n>0} n (z_n
\frac{\partial}{\partial z_n} + \frac{1}{2} )
\label{withzeropoint}
\ee
(Although this expression looks infinite,  we will see that this
expression that also happens to have a well-defined spectrum.)
This is also compatible with the commutation relations of the
Virasoro algebra since there is  an ambiguity that appears when we
evaluate the commutator $[L_1,L_{-1}]$. In it appears the sum
\be
\sum_{n>0} (z_{n+1} \partial_{n+1} - z_n \partial_n) \sqrt{n(n+1)}
\ee
that can be evaluated to give $-\sum_{n>0} n z_n \partial_n$, but
also to give $-\sum_{n>0} n \partial_n z_n$.  The choice of normal
ordering given above amounts to picking the answer ``in the
middle" of these two expressions.

The other positive Virasoro generators
are
\be
L_m =-\sum_{n>0} \sqrt{(m+n)n} z_{m+n}
\frac{\partial}{\partial z_n} + \alpha_0 \sqrt{m} z_m + \frac{1}{2}
\sum_{n=1}^{m-1} \sqrt{n(m-n)} z_n z_{m-n}
\ee
and similar expressions for the negative Virasoro generators. We
conjecture that with these differential operators there is a
unitary irreducible representation of the Euclidean Virasoro
algebra on the space of complex square integrable functions of
$z_k,\bar{z}_k$.

\subsubsection{Intermezzo: A non-positive definite realization of the hermiticity
conditions}

Before studying unitary representations of the Euclidean Virasoro
algebra, we first provide some details of a non-positive definite
realization. Using unitary representations of $SL(2,C)$ is rather
complicated. As above, we can stay closer to the usual intuition
regarding highest and lowest weight states, and sacrifice
unitarity.

Define a pairing $B(|\psi\rangle,|\chi\rangle)$,
which is antilinear in the first and linear in the second
argument, in such a way that
\be
B(|\psi\rangle,L_{k}|\chi\rangle) = B(-\bar{L}_k
|\psi\rangle,|\chi\rangle))
\ee
and similarly for $\bar{L}_k$.   We give up unitarity by not demanding
positive definiteness of this bilinear form. The natural
Hilbert space on which such a pairing acts is generated by a
highest weight state $|h,\bar{h}\rangle$ which is annihilated by
$L_k$, $k>0$ and $\bar{L}_{-k}$, $k>0$. Notice the similarity to
the discussion in \cite{wittencs}. The
pairing can be written in conventional form as follows. Given a
state $\prod_{i,j} L_{-k_i} \bar{L}_{l_j} |h,\bar{h}\rangle$ we
define a bra state
\be
\langle -\bar{h}^{\ast},-h^{\ast} |
\left( \prod_{i,j} L_{-k_i} \bar{L}_{l_j} \right)^{\dagger}
\ee
where the dagger operation is our new hermiticity condition, and
$h^{\ast}$ is the complex conjugate of $h$. The two entries in the
highest weight state denote its $L_0$ and $\bar{L}_0$ eigenvalue.

The inner product of two states is now described as usual by the
product of a bra and a ket state. The latter is completely fixed
once we use the Virasoro algebra, and use that
\be
\langle -\bar{h}^{\ast},-h^{\ast} | h,\bar{h}\rangle =1 .
\ee

This defines a non-positive definite inner product, with respect
to which we do have the new hermiticity conditions. What are the
properties of this inner product? Many states have norm zero.
However, there is a modified notion of positive definiteness.
Consider the ${\bf Z}_2$ involution that sends $L_{k}$ to
$-\bar{L}_{-k}$ and similarly for $\bar{L}_k$. This is not an
anti-involution, like $\dagger$, but a linear map, and it gives us
a ${\bf Z}_2$ involution of the Hilbert space we have defined.
Call this involution $\omega$. Then we have the property
\be
B(\omega |\psi\rangle, |\psi\rangle)=B( |\psi\rangle,
\omega |\psi\rangle)
\geq 0 .
\ee

Therefore, the $+1$ eigenspace of $\omega$ has positive norm, the
$-1$ eigenspace has negative norm. This suggests to project out the
$-1$ eigenspace of $\omega$. However, in this new Hilbert space we
no longer have a representation of the full Virasoro algebra, but
only of the generators that commute with $\omega$, which are
$L_{k}-\bar{L}_{-k}$. The ${\bf Z}_2$ invariant Hilbert space
carries therefore a unitary representation of $SL(2,R)$, but not
of $SL(2,C)$.

This situation recalls the quantization of a free scalar field in
de Sitter space, where solutions to the free wave equation do not
have positive definite Klein-Gordon norm.   Also, the map $\omega$
is reminiscent of the CPT map discussed in \cite{wittds}.

A place where the above discussion could be become especially
relevant is in elliptical de Sitter space, discussed first by
Schr\"odinger \cite{schrodinger}.  Elliptical de Sitter space is
the quotient of de Sitter space by the antipodal map, which maps
$X_i \rightarrow -X_i$ in the definition (\ref{exx1}) of de Sitter
space, resulting in a manifold that is not time orientable.   (Also see \cite{verlindeetal}.)  This $Z_2$ action is exactly like the map $\omega$, and the
$+1$ eigenspace of $\omega$ in the above setup is an interesting
candidate for more conventional CFT description of elliptical de
Sitter space.

\subsubsection{Spectrum of $L_0,\bar{L}_0$}

The zero momentum contributions to $L_0$ and $\bar{L}_0$ read
$(p_1-ip_2)^2/8$ and $(p_2-ip_1)^2/8$, which satisfy
$L_0^{\dagger} = -\bar{L}_0$, as required by hermiticity. To
determine the complete spectrum of $L_0$ and $\bar{L}_0$ is not so
easy.   Consider the subset of the Hilbert space spanned by the
square integrable functions of $z_1,\bar{z}_1$. On these
functions, $L_0$ acts as $-z_1 \partial_1$, and $\bar{L}_0$ acts
as $-\bar{z}_1 \bar{\partial}_1$.
Below we will argue that these operators act on wavefunctions just
like generators of $SL(2,C)$ acting on a principal series
representation.   We will use this to determine the spectrum of
our free scalar with unusual Hermiticity conditions and, in the
next section, use this to explore the partition function.

First, we briefly review the principal series representations of
$SL(2,C)$. These are realized on functions of a complex variable
$f(z)$, and $SL(2,C)$ acts as
\be \label{slrep}
T_g f(z) = \frac{1}{(cz+d)^{\nu_1} (\bar{c} \bar{z} + \bar{d})^{\nu_2}}
f(\frac{az+b}{cz+d}).
\ee
The inner product is given by
\be
\langle f,g\rangle = \int d^2 z \bar{f(z)} g(z) .
\ee
Strictly speaking, the action of $SL(2,C)$ is defined on test
functions on the complex plane that vanish faster than any power
at infinity.   At the end, we take the Hilbert space completion of
the set of test functions. One may check that the inner product is
invariant only if
\be
\nu_1 + \bar{\nu}_2 - 2 =0.
\ee
In addition, the function is only well-defined if
$
\nu_1-\nu_2 \in {\bf Z} .
$

The case with $\nu_1=\nu_2=1+2i\gamma$, $\gamma\in {\bf R}$
corresponds to the usual principal series representations (see
\cite{gaw, teschner9712, gelfand5}).  The simplest case is the one with
$\gamma=0$, so that $\nu_1=\nu_2=1$. Taking $a=1+\epsilon$ and
$d=1-\epsilon$ and expanding to first order in $\epsilon$, we get
\be
f(z) \rightarrow 2\epsilon(\frac{\nu_1}{2} + z\partial) f(z) +
2 \bar{\epsilon} (\frac{\nu_2}{2} + \bar{z} \bar{\partial}) f(z).
\ee
Therefore, in the principal series representation with
$\nu_1=\nu_2=1$  the scaling generators in $SL(2,C)$ act as
 $\CL_0=-(z\partial+1/2)$ and $\bar{\CL}_0=-(\bar{z} \bar{\partial} + 1/2)$, exactly
matching each term in the series in (\ref{withzeropoint})
including the $1/2$. Therefore, it appears that we can compute the
spectrum of $L_0$ for our free boson with novel Hermiticity
conditions by first computing the spectrum of $\CL_0$ and
$\bar\CL_0$ and then summing over principal series
representations.

An eigenfunction of $\CL_0$ with eigenvalue $\lambda$, and with
eigenvalue $\bar{\lambda}$ of $\bar{\CL}_0$, must be
$z^{-\lambda-1/2} \bar{z}^{-\bar{\lambda}-1/2}$.  This function
fails to be integrable at infinity, zero, or both.  Nevertheless,
the operators $\CL_0$ and $\bar\CL_0$ have a well-defined
spectrum\footnote{A number $\lambda$ is in the spectrum of an
operator $A$ is $(A-\lambda)^{-1}$ exists, is bounded, and is
defined on a dense set.} in such a principal series representation
\cite{gaw}. The spectrum is such that $(\CL_0-\bar{\CL}_0)\in {\bf
Z}$ and $i(\CL_0+\bar{\CL}_0)\in {\bf R}$.   The first of these
conditions is natural, since the first operator corresponds to
rotations in two dimensions and therefore generates a $U(1)$.

Armed with this information, and looking back at
(\ref{withzeropoint}) we see that eigenfunctions of
$L_0,\bar{L}_0$ can be taken as a product of functions of the
$z_i$, each of which is an eigenfunction of $\CL_0, \bar{\CL}_0$
in a principal series representation.   Altogether, we find that
the total spectrum of $L_0,\bar{L}_0$ is precisely as it is in a
principal series representation.

\subsubsection{Characters of Euclidean Virasoro}
\label{sec:char}

The partition function of our Euclidean Virasoro algebra is
closely related to the character formula for $SL(2,C)$. In
\cite{bal} the definition of the character of a unitary
representation of locally compact groups is reviewed.    The
result for $SL(2,C)$ is described, for example, in \cite{naimark}.
Take $\nu_1= 1 -m/2-i\rho/2$, $\nu_2=1 + m/2-i\rho/2$. Then the
character reads
\be \label{e12}
{\rm Tr} R \left(
\begin{array}{cc} \lambda & 0 \\ 0 & \lambda^{-1} \end{array}
\right) = \frac{ |\lambda|^{i\rho-m} \lambda^m +
|\lambda|^{-i\rho+m} \lambda^{-m} }{ | \lambda - \lambda^{-1}|^2} .
\ee
Up to signs this is somewhat reminiscent of the usual character
formula for $SU(2)$.
If the representation has $\nu_1=\nu_2=1$, we get simply
\be
{\rm Tr} R \left(
\begin{array}{cc} \lambda & 0 \\ 0 & \lambda^{-1} \end{array}
\right) = \frac{2}{ | \lambda - \lambda^{-1}|^2} .
\ee
Now the matrix appearing in the left hand side is simply the
unitary operation
$
U = e^{-2 \log \lambda \CL_0 - 2 \log \bar{\lambda} \bar{\CL}_0}
$
and we get
\be
{\rm Tr} (q^{\CL_0} \bar{q}^{\bar{\CL}_0}) = \frac{2}{(q^{1/2} - q^{-1/2})(
\bar{q}^{1/2} - \bar{q}^{-1/2}) } .
\ee

This is real, which is consistent with the fact that the operator
appearing here is unitary.  Recall that the $L_0$ and $\bar{L}_0$
operators appearing in (\ref{withzeropoint}) look like an infinite
sum of $\CL_0$ and $\bar{\CL}_0$ operators in principal series
representations of $SL(2,C)$.  Now, using the hermitian operators
$X=i(L_0+\bar{L}_0)$ and $Y=L_0-\bar{L}_0$ we extrapolate to find
the Virasoro character
\be \label{e14}
{\rm Tr}(q_1^X q_2^Y) = \sum_{\alpha_0,\beta_0}
q_1^{i(\alpha_0^2 + \beta_0^2)/2}
q_2^{(\alpha_0^2 - \beta_0^2)/2}
\prod_{k=1}^{\infty}
\frac{2}{((q_1^i q_2)^{k/2} - (q_1^i q_2)^{-k/2})((
q_1^i q_2^{-1})^{k/2} - (q_1^i q_2^{-1})^{-k/2})} .
\ee
From this interesting equation for the Euclidean Virasoro
character two natural questions arise: (a) Can we compute the
density of states from this expression?, (b) Can we build modular
invariants from it? Actually our expression for the Virasoro
character (\ref{e14}) is somewhat formal.  For instance, it
clearly vanishes if $|q_1^i q_2|\neq1$ because the terms in the
product vanish for large $k$. We therefore view (\ref{e14}) as a
formal expression that encodes part of the structure of the
Hilbert space. Notice that in more conventional variables the
above expression is tantalizingly close to the ordinary Virasoro
character
\be
\label{e22}
{\rm
Tr}(q^{L_0} \bar{q}^{\bar{L}_0})= q^{\alpha_0^2/2}
\bar{q}^{\beta_0^2/2}\prod_{k=1}^{\infty}
\frac{2}{(q^{k/2}-q^{-k/2})(\bar{q}^{k/2}-\bar{q}^{-k/2})} \, ,
\ee
but we have formally analytically continued to values where $q$
and $\bar{q}$ are no longer each other's complex conjugate.

\subsubsection{The interpretation of these novel CFT's}

We have gone a long way towards implementing the novel reality
conditions suggested by de Sitter physics in a toy model of a free
scalar.  There are some interesting features, such as an unbounded
spectrum of $L_0$, but perhaps that is what we should expect for a
time-dependent background.

To what extend are these theories ordinary $1+1$-dimensional field
theories? A field theory in $1+1$ dimensions has as local symmetry
group a semidirect product of translations in the $\sigma,t$
directions and $SO(1,1)$ Lorentz rotations. The hermitian
generators are $K=i(t\partial_{\sigma} + \sigma \partial_t)$,
$P_t=i\partial_t$ and $P_{\sigma}= i\partial_{\sigma}$. These form
a simple algebra, the only nonzero commutators are
$[P_t,K]=iP_{\sigma}$ and $[P_{\sigma},K]=iP_{t}$. Can this be a
subalgebra of $sl(2,C)$? Let's consider the basis
$X_k=i(L_k+\bar{L}_ki)$, $Y_k=L_k-\bar{L}_{-k}$ of hermitian
generators. The commutators are
\be
[X_k,X_l]=-[Y_k,Y_l]=i(k-l)X_{k+l}, \qquad [X_k,Y_l]= [Y_k,X_l]=
i(k-l)Y_{k+l}.
\ee
We find that $P_t=X_{-1}$, $P_{\sigma}=Y_{-1}$ and $K=-X_0$
satisfy the right algebra. Therefore, there is an extension of the
symmetry group in $1+1$ Lorentzian dimensions which is $SL(2,C)$.

The novel CFT's that we consider here can therefore be ordinary
$1+1$ dimensional field theories, where the Poincar\'e group is
extended to $SL(2,C)$, rather than $SL(2,R)\times SL(2,R)$. This
is not the same as a Euclidean conformal field theory, which
carries unitary representations of $SL(2,R)\times SL(2,R)$.

One may try to construct a Lagrangian for the toy model by
starting with Lagrangians for chiral bosons, and combining these
in a suitable way. It would be interesting to work this out in
more detail.

\subsection{Application to de Sitter Holography}

Above we have implemented the novel reality conditions suggested
by de Sitter physics in a toy model of a free scalar. We now
explore how CFTs realizing our Euclidean Virasoro algebra would be
expected to represent the physics of de Sitter space.

As we have seen in earlier sections, there are many reasons to
believe that two CFT's, rather than one, are relevant. De Sitter
has two boundaries rather than one (although they are not
completely independent), and in the static patch we see a thermal
spectrum which typically arises in situations where one entangled
factor in a product of two field theories is integrated out.
Indeed, the thermal state of free fields in the static patch can
be understood in terms of integrating out the correlated region in
the antipodal patch.  The argument that a dual to 3d de Sitter
would involve only one CFT \cite{andyds} is partly based on the
fact that in global coordinates there is an $SL(2,C)$ invariant
vacuum \cite{vacuum}, which is what we have in a single
CFT\footnote{In the toy model, however, there is no obvious
$SL(2,C)$ invariant `vacuum state'.}. Likewise, the isometry group
of 3d de Sitter is a single $SL(2,C)$ which again suggests one
dual CFT, not two.

However, there is a potential way around this dilemma. It is
possible that de Sitter is described by a pure, but entangled
state in two CFTs which preserves a single $SL(2,C)$ and that
tracing over a suitable part of the Hilbert space yields a thermal
density matrix.\footnote{This would be reminiscent of the
situation in the eternal black BTZ black hole \cite{lorentzian2,
eternalbh}.}    In conventional tensor products of CFT's, the only
$SL(2,C)$ invariant state is the product of the vacua of the two
theories. The tensor product of two highest weight states is not
invariant under $1\otimes L_0 + L_0 \otimes 1$.

Once we change the hermiticity conditions, the situation changes
completely. We can quite generally construct $SL(2,C)$ invariant
states, using the unitary representations of section~3.1.1, but
also using the unitary but not positive definite realization of
section~3.1.4.

Below we discuss these possibilities for realizing an $SL(2,C)$
invariant state in a product of two CFTs associated with the early
and late time de Sitter boundaries. Next, we give a concrete
example of an $SL(2,C)$-invariant, pure state in the tensor
product of two of our toy models. We then study how quite
generally the Hilbert space could be split up in terms of Hilbert
spaces associated with antipodal static patches, realizing a kind
of de Sitter complementarity.  We conclude by discussing the
picture of de Sitter duality emerging from our explorations.

\subsubsection{Pure states with $SL(2,C)$ Invariance}

\paragraph{Using the novel hermiticity conditions:}
Consider the unitary $SL(2,C)$ representation with
$\nu_1=\nu_2=1$. As discussed above, this representation appears
in the toy model, and it can be extended to a representation of
the Virasoro algebra with the new hermiticity conditions. If we
call this representation $C$, then $C\otimes C$ contains the
identity representation, which is crucially different from the
case where we take two discrete highest weight representations of
$SL(2,R)$. We can see this since there is a map from $C \otimes C$
to the complex numbers (which form the trivial representation of
$SL(2,C)$):
\be
\label{e15}
(\phi_1(z),\phi_2(z')) \longrightarrow \int d^2 z \phi_1(z)
\phi_2(z) \, .
\ee
We also could have chosen
\be
\label{e16}
(\phi_1(z),\phi_2(z')) \longrightarrow \int
\frac{d^2 z d^2z'}{|z-z'|^2} \phi_1(z) \phi_2(z') \, .
\ee
One can check that both these expressions are invariant under the
action of $SL(2,C)$.   The second one will not play any major role
here but would be important if we were not using the
representation with $\nu_1=\nu_2=1$. Using this pairing, we can
make a pure state in the tensor product of two Euclidean Virasoro
representations, that is invariant under $SL(2,C)$. We simply take
the dual map of (\ref{e15}) or (\ref{e16}), and consider the image
of an arbitrary nonzero complex number under the dual map.

An explicit example of such a pure, entangled state will be given
in the following subsection.

\paragraph{Non-unitary pure states with $SL(2,C)$ invariance}
We can  also describe the pure states that appear when we use as
Hilbert space a combination of highest and lowest weight
representations. The construction proceeds exactly as above, but
we do not use the fact the tensor product of two principal series
representations contains the identity. Instead we observe that the
tensor product of a highest and a lowest weight representation
with highest and lowest weight $+h$ and $-h$ respectively,
contains the identity representation. It is slightly cumbersome to
describe this pure state. First, we introduce another ${\bf Z}_2$
involution, called $I$, that sends $L_{k} \rightarrow -L_{-k}$ and
similarly for $\bar{L}_{k}$.   In contrast to the dagger
operation, this is a linear, not an antilinear operation. We also
need the standard inner product which we know from string theory,
and denote that by a pairing $B_{\rm
standard}(|\psi\rangle,|\chi\rangle)$. Let $|\psi_{\alpha}\rangle$
be a basis for the module built on $|h,\bar{h}\rangle$. Then
$I|\psi_{\alpha}\rangle$ is a basis for the module built on
$|-h,-\bar{h}\rangle$. We also define $M_{\alpha\beta}= B_{\rm
standard}(|\psi\rangle_{\alpha},|\psi_{\beta}\rangle)$, and its
inverse $M^{\alpha\beta}$. Then the pure state is
\be
\sum_{\alpha,\beta} M^{\alpha\beta} |\psi_{\alpha}\rangle \otimes
I |\psi_{\beta}\rangle
\ee
and one easily verifies it preserves the diagonal Virasoro
subalgebra with $L_m$, $m\geq -1$, and $\bar{L}_m$, $m\leq +1$.
While this construction gives an SL(2,C) invariant state, it does
not arise from a unitary implementation of SL(2,C). Again, this
state may be relevant for a dual description of the $Z_2$ quotient
of de Sitter space by the antipodal map, as discussed in
section~3.1.4.

\subsection{An example of a pure entangled state}

To describe an example of a pure entangled state with $SL(2,C)$
invariance, we take two copies of the toy model. The modes of the
two models are denoted by
$\alpha^1_n,\beta^1_n,\alpha^2_n,\beta^2_n$. We introduce the
following two linear combinations
\bea \label{newmodes}
U_n & = & \frac{1}{2}( \alpha^1_n + i \beta^1_{-n} + i \alpha^2_n
-\beta^2_{-n}) \nonumber \\
V_n & = & \frac{1}{2}(\alpha^1_n - i \beta^1_{-n} + i \alpha^2_n +
\beta^2_{-n}) .
\eea
The commutations relations between these operators are
\be
[U_n,U^{\dagger}_{m}]=-[V_n,V^{\dagger}_{m}]=n\delta_{n-m,0}.
\ee
We will call $U,V$ annihilation operators and
$U^{\dagger},V^{\dagger}$ creation operators. The nice feature of
these operators is that $L^1_{k}+L^2_{k}$ and $\bar{L}^1_k +
\bar{L}^2_k$ will only consist of terms that contain one creation
operator and one annihilation operator, but no terms with two
creation or two annihilation operators. To show this, we take a
typical term $\alpha^1_k \alpha^1_l + \alpha^2_k \alpha^2_l$ in
$L^1_{k+l} + L^2_{k+l}$. Using
\bea
\alpha^1_n & = &  \frac{1}{2}(U_n + U^{\dagger}_{-n} + V_n
-V^{\dagger}_{-n}) \nonumber \\
\alpha^2_n & = &  \frac{1}{2}(-i U_n + i U^{\dagger}_{-n} -i V_n
-i V^{\dagger}_{-n})
\eea
we get
\be
2(\alpha^1_k \alpha^1_l + \alpha^2_k \alpha^2_l)  =
(U_k+V_k)(U^{\dagger}_{-l} - V^{\dagger}_{-l}) +
(U^{\dagger}_{-k}-V^{\dagger}_{-k})(U_l + V_l)
\ee
and
\be
2(\beta^1_k \beta^1_l + \beta^2_k \beta^2_l)  =
(U_k-V_k)(U^{\dagger}_{-l} + V^{\dagger}_{-l}) +
(U^{\dagger}_{-k}+V^{\dagger}_{-k})(U_l - V_l)
\ee
Because the diagonal $SL(2,C)$, spanned by $L^1_{k}+L^2_{k}$ and
$\bar{L}^1_k + \bar{L}^2_k$ with $k=-1,0,1$, contains only terms
linear in creation and annihilation operators, an entangled
$SL(2,C)$ invariant state is the $|0\rangle>$ that satisfies
\be U_n|0\rangle = V_n |0\rangle =0 .
\ee

This way of constructing an $SL(2,C)$ invariant state is quite
reminiscent of the construction of such states in the Hilbert
space of a free scalar field on de Sitter space. In that case, the
$SL(2,C)$ generators are also sums of products of creation and
annihilation operators, as we will explain in more detail in
section~4.

If we consider excitations of the state $|0\rangle$, we find excitations
with negative norm. This is because the true annihilation operators
are $U_n,U^{\dagger}_{-n},V_{-n},V^{\dagger}_n$ with $n\geq 0$.
Therefore, $|0\rangle$ is not a really good candidate 
$SL(2,C)$ invariant state. We can do better by starting with a state
$|\Omega\rangle$ which is annihilated by the true annihilation operators.
Then we can construct a squeezed state
\be
|{\rm entangled}\rangle = \exp \left[
\sum_{k>0} \frac{1}{k} ( U^{\dagger}_k V_k + U_{-k} V^{\dagger}_{-k} )
\right] |\Omega\rangle 
\ee
which is $SL(2,C)$ invariant, as one may verify. 

If one tries to construct an $SL(2,C)$ invariant state for a
single copy of the toy model, one finds that similar tricks do
not work. This we view as an encouraging sign towards the
correctness of the holographic dual picture of de Sitter space we
are proposing here. It is also worth pointing out that the above
construction bears in many ways a close similarity to the
construction of boundary states. This analogy may be useful when
trying to find $SL(2,C)$ invariant states in more complicated
examples.

With this explicit example of an $SL(2,C)$ invariant state in
hand, it is in principle possible to start doing explicit
calculations of correlations functions in the product of two toy
models, and to compare those with bulk calculations; we leave this
to future work, as well as a detailed analysis of the normal
ordering issues that appear in this construction.

\subsubsection{Static versus global patch and de Sitter thermodynamics}

It is a well-known fact that the static observer in de Sitter
space sees a thermal spectrum of excitations when de Sitter is
placed in the Euclidean vacuum state.  One way of understanding
this thermality for a scalar field in the static patch is by
integrating out the fluctuations in the antipodal static patch.
The static patch of de Sitter space also has a horizon giving rise
to gravitational entropy.   Integrating out quantum fields in the
antipodal patch cannot quite account for this entropy because of
the usual ultraviolet problems afflicting such ``geometric
entropies''.   Instead, in previous explorations holography, 3d de
Sitter entropy was associated with the degeneracy of particular
states in a highest weight representation of a dual CFT.   For the
Kerr-de Sitter spaces with ``mass" $M$ and angular momentum $J$,
it was found that these states would have $L_0 \sim M + i J$ and
similarly for $\bar{L}_0$, where $L_0, \bar{L}_0$ are generators
of time translation in the static patch but are radial
translations in the plane at the late time de Sitter boundary.
Applying the Cardy formula naively to such states yielded the de
Sitter entropy \cite{bdbmds, strombousso, park, banados}.

Here we explore how the physics of  antipodal static patches could
be implemented in a holographic setting where a dual to de Sitter
involves two CFTs, one associated with each of the early and late
time boundaries.   Denote the Hilbert spaces of these two theories
as ${\cal H}_{i,f}$.    We assume that both ${\cal H}_{i,f}$ are
CFT's based on the Euclidean Virasoro algebra, since $SL(2,C)$
acts separately at plus and minus infinity.   Since global de
Sitter space has a single $SL(2,C)$ isometry group, we expect it
to be represented in the dual theory as an entangled state with an
$SL(2,C)$ invariance, perhaps constructed  as described above.

How can we construct Hilbert spaces ${\cal H}_{L,R}$ associated
with two antipodal static patches from ${\cal H}_{i,f}$?   From
the action of $SL(2,C)$ on the bulk of de Sitter space, we know
that neither of ${\cal H}_{L,R}$ should carry an action of
$SL(2,C)$ by itself.   It is interesting to learn how this mixing
works.  For that purpose, consider a  $U(1)$ that rotates by an
angle $\theta$ along the equator of the sphere at equal time
slices in global de Sitter.   Since the isometry group is really
$SL(2,C)/{\bf Z}_2$, the rotation is implemented by
 \be
U(\theta) =\left(
\begin{array}{cc}
\cos \theta/2 & \sin \theta/2 \\
 -\sin \theta/2 & \cos \theta/2
\end{array} \right)
\ee
On the 2-sphere boundary of de Sitter at early or late times, this
rotation is the map $z\rightarrow -1/z$, which is quite similar to
the CPT map of Witten \cite{wittds}.   One can easily see how it
acts on e.g. $L_0$, by computing $U(\theta) L_0 U(\theta)^{-1} $.
One gets
\be
U(\theta) L_0 U(\theta)^{-1} = \cos (2\theta) L_0
-\sin(2 \theta) (L_{-1}+L_1).
\ee
Thus, for a rotation through $\pi$, $L_0$ flips signs.

Recall from above that the holographic discussions of de Sitter
entropy \cite{bdbmds, strombousso, park, banados} suggest that
e.g. ${\cal H}_L$ is built out of standard highest weight modules.
(It was using this assumption that the entropy of the static patch
was reproduced from a CFT dual to de Sitter.)    The fact that
$L_0$ flips sign in going between two antipodal static patches
then naturally means that ${\cal H}_R$ is built out of lowest
weight modules.  To get some sense of the relation between the
Hilbert spaces $\CH_{i,f}$ based on the Euclidean Virasoro algebra
and $\CH_{L,R}$ we should compare the partition function in the
former case with the partition function in a highest/lowest weight
representation.  In the absence of a concrete proposal for a CFT
dual to de Sitter space we again explore the free scalar field
from Sec.~\ref{sec:toymodel}.   The trace of $q^{L_0}
\bar{q}^{\bar{L}_0}$ in $\CH_{i,f}$ is given in (\ref{e22}) while
the trace in highest/lowest weight representations is well known:
\bea
\chi_i = \chi_f & \sim &
\prod_{k>0} \frac{1}{|q^{k/2}-q^{-k/2}|^2} \nonumber \\
\chi_L & \sim & \prod_{k>0} \frac{1}{|1-q^k|^2} \nonumber \\
\chi_R & \sim & \prod_{k>0} \frac{1}{|1-q^{-k}|^2},
\eea
where $\chi$ denotes the trace of $q^{L_0} \bar{q}^{\bar{L}_0}$
over the relevant Hilbert space (with the naive inner products).
(We have left out numerical factors and zero modes for the moment.)

Interestingly,
\be
\chi_i \chi_f = \chi_R \chi_L
\ee
which in turn suggests that
\be
{\cal H}_f \otimes {\cal H}_i \equiv
{\cal H}_L \otimes {\cal H}_R .
\ee
This suggests that there are two different ways of thinking about
the same Hilbert space: we could either regard it as the Hilbert
space of two CFTs defined at past and future infinity of de Sitter
space realizing Euclidean Virasoro algebras, or as the Hilbert
space of antipodal static patches based on lowest/highest weight
representations\footnote{Incidentally, yet another point of view
is that the rotation above does not really map a state into a
state, but rather it maps a particle into an antiparticle. In that
case, one should view ${\cal H}_L$ as the bra states, and ${\cal
H}_R$ as the ket states, and the $U(1)$ rotation mixes bra and ket
states. This is again somewhat similar to \cite{wittds}. 't Hooft
\cite{bhthooft} has also discussed a similar idea in the context
of black holes; in this case bra and ket states were associated to
different regions.}.   The two radically different bases would
have a complicated, possibly non-local relationship to each other,
perhaps realizing a sort of horizon complementarity \cite{compl}
in de Sitter space.   We will explore this further that the level
of free fields in de Sitter space in
Sec.~\ref{sec:freescalarevid}.

\subsubsection{A picture of holographic duality in de Sitter}

In summary, our picture is that the $SL(2,C)$ invariant vacua of
de Sitter space should be realized as entangled states in ${\cal
H}_L \otimes {\cal H}_R$, with the thermally entangled state
corresponding to the Euclidean vacuum. This entanglement would
account for de Sitter when entropy when one of the factors is
traced over.    Other entanglements would give rise to the
one-parameter family of de Sitter vacua.\footnote{It is possible that once interactions are taken into account all vacua except the Euclidean one will be eliminated.}   All of these entangled
states should translate into $SL(2,C)$-invariant states in ${\cal
H}_i \otimes {\cal H}_f$, on which the Euclidean Virasoro symmetry
acts in a simple way.  It would be very interesting to firmly
establish that such a picture, including the one-parameter family
of vacua, is realized at least in a toy model.

This picture does not  contradict the picture of de
Sitter space given by Witten in \cite{wittds}. In that paper,
the claim is that all quantum gravity can do is to provide an
Hilbert space structure for de Sitter space. A connection with our proposal can be
made, once we identify the set of all correlation functions of one
of the two Euclidean CFT's with some large vector space (in the
spirit of the Chern-Simons - conformal field theory relation). The
Hilbert space structure then arises from the entangled state.
Given two correlation functions in each of the CFT's, the Hilbert
space structure is then given by computing the joint correlation
function in the entangled state that links the two CFT's.

\section{Further evidence: free scalar field}
\label{sec:freescalarevid}

A great deal was learned about the AdS/CFT correspondence by studying the physics of free scalar fields in AdS space.  Therefore, in this section we study the massive, free scalar field in de Sitter, in order
to collect further evidence for the picture of dS holography that we
have given above.  Free, massive scalars in de Sitter space have been
extensively studied in the literature (see, e.g, \cite{vacuum}, and the recent works \cite{strombousso,volosprad}).  Here, we will focus on the
representations that the Hilbert space of a free scalar field form
under the action of the isometry group.   The representation theory corroborates the picture of de Sitter holography developed above.  Along the way
we will recover many known results and also find several new ones.

First, we will review why isometries are implemented unitarily on
the Hilbert space of a canonically quantized scalar field. Next, we study the Hilbert space of the scalar
field, and show that the one-particle states built from an
$SO(d,1)$-invariant vacuum state form a principal series
representation of $SO(d,1)$ (sometimes called class 1
representations). Next, we determine what form states have that
have a fixed eigenvalue under $L_0+\bar{L}_0$ (or its
generalization to $SO(d,1)$. When restricted to the static patch,
these are states with a fixed frequency $\omega$. We also find a
basis of states in which the decomposition of the Hilbert space in
modes living on the two separate static patches becomes
particularly simple. This latter decomposition also provides a
clue regarding which states we should trace over in a holographic dual
description, in order to find the density matrix that describes
one static patch. Finally, we briefly consider the particularly
simple case of a massless scalar in $d=2$.

\subsection{Isometries}

First we reexamine the way isometries of a space are
implemented on the Hilbert space of a canonically quantized scalar
field.  Consider a space with metric (for simplicity)
\be
ds^2 = -N^2 dt^2 + g_{ij} dx^i dx^j .
\ee
If this space has an isometry
$
\delta t =\lambda,
\delta x^i = \xi^i ,
$
we have the usual equations
\bea
\lambda \partial_t N + \xi^i \partial_i N + N\partial_t \lambda &
= & 0 \nonumber , \\
-N^2 \partial_i \lambda + g_{ij} \partial_t \xi^j & = & 0
\nonumber , \\
\lambda \partial_t g_{ij} + \xi^k \partial_k g_{ij}
+ \partial_i \xi^k g_{kj} + \partial_j \xi^k g_{ki} & = & 0 .
\eea
The field equation of a free scalar reads
\be
\partial_t (N^{-1} \sqrt{g} \partial_t) \phi -
\partial_i (N \sqrt{g} g^{ij} \partial_i) \phi
+ m^2 N \sqrt{g} \phi = 0  \, ,
\ee
with a Klein-Gordon norm,
$
\langle \phi,\chi \rangle =
-i \int dx \sqrt{g} N^{-1} (\phi \dot{\chi}^{\ast} -
\dot{\phi} \chi^{\ast} )  ,
$
which is time-independent
($
\partial_t \langle \phi,\chi \rangle =0
$),
and invariant under the isometries
($
\langle \delta \phi, \chi \rangle +
\langle \phi, \delta \chi \rangle =0 
$).
To prove the latter statement, one can do an explicit calculation,
using the equations of motion of the scalar field and the various
equations for an isometry.  The fact that isometries leave the
Klein-Gordon norm invariant suggests that isometries give rise to
anti-hermitian operators in the quantum theory, so that the
isometry group is unitarily implemented. To verify this, we have
to understand in more detail the relation between the Klein-Gordon
norm and canonical quantization.

To do canonical quantization, we first write the most general
solution of the equations of motion
\be \label{f1}
\phi = \sum a_m u_m(x,t)
\ee
where $u_m$ includes all modes, including positive and negative
frequency modes. In general, $u_m^{\ast}$ is also a set of
solutions, and can therefore be expressed in terms of $u_m$,
\be \label{f2}
u_m^{\ast} = \sum_n H_{mn} u_n .
\ee
The canonical momentum
$
\pi=\sqrt{g} N^{-1} \partial_t \phi
$
satisfies
$
[\phi(t,x),\pi(t,x')]=i \delta(x-x'),
$
and in addition the field and momentum commute with themselves.
Inserting the expansion (\ref{f1}) we find that
\bea
\sum_{m,n} [a_m,a_n] \sqrt{g}(x',t) N^{-1}(x',t)
u_m(x,t) \dot{u}_n(x',t) & = & i
\delta(x-x') \nonumber \\
\sum_{m,n} [a_m,a_n]u_m(x,t) u_n(x',t) & = & 0 \nonumber \\
\sum_{m,n} \sqrt{g}(x,t) N^{-1}(x,t) \sqrt{g}(x',t) N^{-1}(x',t)
[a_m,a_n] \dot{u}_m(x,t) \dot{u}_n(x,t) & = & 0 .
\eea

We can integrate these identities against $u_r$ and $\dot{u}_r$
to get various identities involving the object
\be
L_{mn} = \int dx \sqrt{g} N^{-1} u_m \dot{u}_n
\ee
in terms of which the commutator is very simple
\be
[a_m,a_n]=-i(L_{mn}-L_{nm})^{-1} .
\ee
Thus, the commutator is expressed directly in terms of the
Klein-Gordon norm
\be
L_{mn}-L_{nm}= i \langle u_m,u_n^{\ast} \rangle_{KG}
\ee
and together with the hermiticity conditions in (\ref{f2}) that
imply
\be
\sum_m a_m^{\dagger} H_{mn} = a_n
\ee
we see that all canonical structure is encoded in the Klein-Gordon
norm.  To canonically quantize a field  we seek a representation of the
commutation relations plus hermiticity conditions. The
Klein-Gordon norm  may or may not be the inner product on the resulting Hilbert space.

The fact that isometries preserved the KG norm implies that they
preserve the canonical commutation relations. In some sense, this result
should have been expected. We already knew that isometries, via
Noether theorem, give rise to a conserved charge that becomes an
operator. They are thus canonical transformations, generated by
commuting with the appropriate conserved charge. Canonical
transformations always preserve the commutation relations.

What about unitarity? Since isometries are real, they commute with
the operation of complex conjugation. Therefore, if $Q$ is the
operator that implements the isometry, then $[Q,a^{\dagger}]=
[Q,a]^{\dagger}$. This shows that $Q$ is anti-hermitian. Hence,
we have showed  that isometries give rise to a canonical
transformation, with time-independent coefficients, generated by
an anti-hermitian operator. Therefore, any standard quantization of the
commutation relations and the hermiticity conditions will
automatically give a unitary representation of the isometry group.

Incidentally, in Witten's approach to de Sitter space \cite{wittds} we take an
inner product on the Hilbert space that is not compatible with the
hermiticity condition in (\ref{f2}), but one that follows from the
path integral. If the naive inner product was $\langle \cdot,\cdot
\rangle_{\rm standard\,\,\,bulk}$, then
\be
\langle \phi,\chi \rangle_{\rm witten}=
\langle \omega(\phi) ,\chi \rangle_{\rm standard\,\,\,bulk} .
\ee
Here, $\omega$ represents the antipodal map on the sphere.  In other words, combining the anitpodal map with the standard inner product in the bulk of de Sitter gives the usual Hermiticity conditions in a dual Euclidean CFT.   Therefore, in Witten's inner product we
find a unitary representation of $SO(2,d-1)$ rather than
$SO(1,d)$. However, the thermal spectrum in \cite{strombousso}
was computed using the standard quantization rather than Witten's
inner product. This provides additional evidence for the $SL(2,C)$
structure that we are advocating.  Another interesting perspective on Witten's inner product has been provided in \cite{strombousso}.

\subsection{The one-particle Hilbert space of a massive scalar}

Before looking more closely at the massive scalar field, we first
give some expressions for the $SO(d,1)$ generators in terms of the
global coordinates on de Sitter space. Recall that de Sitter space
is described by the equations
\be -X_0^2 +X_1^2 +\ldots + X_d^2 =1 \ee
in a space with metric with signature $(-,+,\ldots,+)$. The entire
space can be parametrized by
\be \label{newcc} X_0 = y \sinh t, \quad X_i = y_i \cosh t \,\,\, \mbox{\rm for}
\,\,\, i>0 \ee with $y^2 = y_1^2 + \ldots y_d^2$ and de Sitter
space is obtained by restricting to $y=1$. Normally, we would
choose spherical coordinates on the $d-1$-sphere parametrized by
$y_i$ with $y=1$, but here we find it more convenient to work with
$y_i$ instead, keeping in mind that $y=1$.

The generators of $SO(d,1)$ are given by
\be M_{ij}= X_i \frac{\partial}{\partial X_j} -
X_j \frac{\partial}{\partial X_i},\qquad K_i = X_0
\frac{\partial}{\partial X_i} - X_i \frac{\partial}{\partial X_0}.
\ee
The generators $M_{ij}$ are obviously the rotations in $d$
dimensions, whereas $K_i$ are boosts. When expressed in terms of
the coordinates (\ref{newcc}), the generators become
\bea \label{gencc}
M_{ij} & =& y_i \frac{\partial}{\partial y_j} - y_j
\frac{\partial}{\partial y_i} \nonumber \\
K_i & = & \frac{y_i}{y} \frac{\partial}{\partial t}
 + \tanh t \left(
 y \frac{\partial}{\partial y_i} - \frac{y_i}{y}
 \sum_k y_k \frac{\partial}{\partial y_k} \right) .
\eea
One may check that these generators leave $y$ invariant, and
therefore reduce to the generators of $SO(d,1)$ acting on de
Sitter space, once we replace $y_i$ by spherical coordinates.

Consider now a massive scalar field on de Sitter space with mass
$m$. A convenient basis of solutions of the equations of motion
are for example the so-called Euclidean modes and their complex conjugates.
An explicit expression can be found in eq. (3.36) in
\cite{strombousso}. Euclidean modes are characterized
by the fact that they admit a regular analytic continuation to the
lower half of the $d$-sphere that represents Euclidean de Sitter
space.  The complex conjugates of the Euclidean modes are regular in the upper half-sphere.
These properties are preserved by acting with isometries, and
therefore the isometries map Euclidean modes into a linear
combination of the Euclidean modes, not involving their complex
conjugates. Therefore, the Euclidean modes furnish a
representation of $SO(d,1)$. What does this representation look
like? The Euclidean modes are in one-to-one correspondence with
elements of $L^2(S^{d-1})$. Indeed, the properly normalized
Eulidean modes are of the form $f_k(t) Y_k(\Omega)$, where $Y_k$
is a spherical harmonic for $S^{d-1}$.

We can therefore set up a one-to-one correspondence between
$L^2(S^{d-1})$ and the Euclidean modes by assigning to $\sum c_k
Y_k(\Omega)$ the Euclidean mode $\sum c_k f_k(t) Y_k(\omega)$. If
the Euclidean modes are properly normalized, this correspondence
maps the standard inner product on $L^2(S^{d-1})$ to the
Klein-Gordon inner product of the Euclidean modes, so it really is
an isomorphism of Hilbert spaces. We argued that the Euclidean
modes formed a representation of $SO(d,1)$, and we therefore also
must find a unitary representation of $SO(d,1)$ on $L^2(S^{d-1})$,
once we use this correspondence.

What is the explicit form of the generators (\ref{gencc}) when
acting on $L^2(S^{d-1})$? The answer is that they become
\bea \label{gencc2}
M_{ij} & =& y_i \frac{\partial}{\partial y_j} - y_j
\frac{\partial}{\partial y_i} \nonumber \\
K_i & = & -2 h^+ \frac{y_i}{y}
 + \left(
 y \frac{\partial}{\partial y_i} - \frac{y_i}{y}
 \sum_k y_k \frac{\partial}{\partial y_k} \right) .
\eea
Here $h^+$ is the scaling dimension given in (\ref{scaldim}),
$2h^+=  (d-1)/2 + i \mu$, with $\mu=\sqrt{m^2 - (d-1)^2/4}$.
To prove (\ref{gencc2}), we first observe that by general
arguments, the generators $M_{ij}$ will remain the same, and that
$K_i$ has to be of the form $$ c\frac{y_i}{y}
 + \left(
 y \frac{\partial}{\partial y_i} - \frac{y_i}{y}
 \sum_k y_k \frac{\partial}{\partial y_k} \right)$$ for some
 constant $c$. The constant $c$ can be determined, for example,
 by acting on the Euclidean mode corresponding to the constant
 spherical harmonic. After some algebra, and using the recursion
 relation
\be
{}_2F_1(a,b,2a,z) + (z-1) {}_2F_1(a+1,b+1,2a+1,z)
=\frac{2a-b}{2(2a+1)} z {}_2 F_1(a+1,b+1,2a+2,z)
\ee
one obtains $c=-2h^+$.

One may for instance compare (\ref{gencc2}) with $d=3$ to the
principal series representations given in (\ref{slrep}). We find
that (\ref{gencc}) is an infinitesimal version of  (\ref{slrep}) with $\nu_1=\nu_2=2h^+$.
For general $d$, the generators (\ref{gencc2}) generate a principal series (or class one)
representation of $SO(d,1)$ on $L^2(S^{d-1})$. It is
straightforward to check that this representation is unitary.

Similarly, the complex conjugates of the Euclidean modes transform
under a principal series representation, with $2h^+$ replaced by
$2h^-$ in (\ref{gencc2}). Since the Euclidean vacuum is
annihilated by the operators that multiply the Euclidean modes,
and the one particle states are obtained by acting with their
hermitian conjugates, the one-particle states made from this
vacuum transform exactly as the complex conjugate Euclidean modes.
Therefore, the one-particle states form a principal series
representation of $SL(2,C)$. This proves statements made in earlier sections, and is further evidence in favor of the new hermiticity conditions we have proposed.

Notice that $SL(2,C)$ maps annihilation operators into
annihilation operators, and creation operators into creation
operators. Therefore, the generators can be written as sums of
products of creation and annihilation operators, and therefore the
state annihilated by all annihilation operators (the Euclidean
vacuum) is indeed $SL(2,C)$ invariant. This structure is nicely
mimicked in our example of an $SL(2,C)$ invariant state in the
product of two toy models in section~3.2.2.

To give an example how $SO(d,1)$ acts on $S^{d-1}$, consider
$d=2$. Then $SO(2,1)=SL(2,R)$, and $SL(2,R)$ acts on a circle via
\be
e^{i\theta} \rightarrow \frac{e^{i(\alpha+\theta)} \cosh\xi +
e^{i\beta} \sinh\xi}{e^{i(\theta-\beta)} \sinh \xi + e^{-i\alpha}
\cosh\xi }
\ee
where $\xi,\alpha,\beta$ are real numbers parametrizing $SL(2,R)$.

The description of the Hilbert space in terms of functions on
$S^{d-1}$ is clearly in the right direction of holography, since
all reference to the time dependence has been lost.   Usually holographic descriptions lose spatial directions, but here time has been eliminated.   This is in keeping with the proposal in \cite{bdbmds} and \cite{rgstrom} (also see \cite{rgothers}) that time evolution in a space with a positive cosmological constant is related to inverse RG flow in a dual Euclidean field theory.

\subsection{Eigenfunctions of a boost}

In the static patch, time translations are clearly an isometry.
This isometry corresponds to a boost generator of $SO(d,1)$. There
are different choices of static patch, and we will restrict our
attention to the static patch where time translation is given by
the boost $K_1$. Solutions of the equation of motion that in the
static patch behave as $e^{i\omega t}$ therefore obey
\be K_1 f = i \omega f . \ee
If we use the correspondence between Euclidean modes and
$L^2(S^{d-1})$, corresponding to $f$ there should be a function
$\psi_{\omega}$ on the sphere, that has the property that
\be
K_1 \psi_{\omega} = i \omega \psi_{\omega} .
\ee
One can readily solve this differential equation and one finds
\be \label{eig}
\psi_{\omega} = \left( \frac{1+y_1}{1-y_1} \right)^{i\omega/2}
(1-y_1)^{-h^+} \chi(\Omega')
\ee
where $\chi(\Omega_{d-2})$ is some function on the unit
$d-2$-sphere in the $y^2\ldots y^d$ directions; it can be
represented by a homogeneous function of degree zero of
$y_2,\ldots,y_d$.

If one were to solve the laplace equation on the $d-1$-sphere, and
would write the metric as $ds^2  = d\theta^2 + \sin^2 \theta
d\Omega_{d-2}^2$ (so that $y_1=\cos\theta$), one also gets a
decomposition of the form $h(y_1) \chi(\Omega_{d-2})$, but now
$h(y_1)$ will typically be a Gegenbauer polynomial. Thus, the
frequency eigenmodes (\ref{eig}) form a completely different basis
compared to the usual spherical harmonics.

To summarize the meaning of (\ref{eig}), suppose we would
decompose (\ref{eig}) in spherical harmonics $\sum c_k Y_k$ on
$S^{d-1}$, and suppose we would write down the corresponding
Euclidean mode $\sum c_k f_k(t) Y_k$. This Euclidean mode, when
restricted to the static patch, will depend on static time $t_s$
as $e^{it_s\omega}$. In particular, the Bogoliubov transformation
that relates modes on the static patch to global modes can be
extracted from the decomposition of (\ref{eig}) in spherical
harmonics. 

It is a straightforward exercise to show that
\be \int dy_1 dy_2 \ldots dy_d \delta(y-1)
\psi_{\omega}(y_i) \bar{\psi_{\omega'}}(y_i) \sim
\delta(\omega-\omega')
\ee
so that the boost eigenfunctions really behave as Fourier modes on
the sphere. They are not really well-defined functions on the
sphere, due to their behavior near $y=\pm 1$. This does not come as
a surprise, since we already noticed before that $L_0$-eigenfunctions
did not exist in the toy model of section~3.

So far, we restricted attention to Euclidean modes. However, there
is a one-parameter family of $SO(d,1)$ invariant vacua, obtained
by taking linear combinations of creation and annihilation
operators. A priori, it may seem strange that this can be done,
since the Euclidean modes transform as (\ref{gencc2}), and their
complex conjugates as (\ref{gencc2}) with $h^+$ replaced by $h^-$.
How can a linear combination of Euclidean modes and their complex
conjugates still transform nicely under $SO(d,1)$? The point is
that the representation (\ref{gencc2}) and the one with $h^+$
replaced by $h^-$ are equivalent representations. In particular,
if $g(y)$ transforms as (\ref{gencc2}) with $h^+$ replaced by
$h^-$, then
\be
{\cal T}(g)(y) \equiv \int d^d z \delta(z-1) (1-\sum_k y_k
z_k)^{-2 h^+} g(z)
\ee
transforms precisely as in (\ref{gencc2}). The integration kernel
of this transform is precisely like a CFT two-point function of two
operators with weight $h^+$.  Therefore, one might expect to make a
more precise translation between properties of functions on
$S^{d-1}$ and properties of a putative holographically dual CFT \cite{andyds,volosprad,strombousso}.

Two particular examples of different basis are to use the in and
out modes discussed for example in section~2.2. These are related
to the Euclidean modes in a simple way,
\bea \label{trafo}
\phi^E & \sim & \phi^{\rm in} + e^{2 \pi i h^+} (\phi^{\rm
in})^{\ast} \nonumber \\
\phi^E & \sim & \phi^{\rm out} + e^{-2 \pi i h^-} (\phi^{\rm
out})^{\ast}.
\eea
In particular, in odd dimensions $e^{2\pi i (h^+ + h^-)}=1$, and
the in and out modes are the same.  Interestingly, the coefficients that appear in
(\ref{trafo}) are similar to terms that appear in
(\ref{eig}) if we try to analytically continue $y_1$ to values
less than $-1$ or greater than $+1$.   Perhaps
all the different choices of modes and vacua in de Sitter space can 
be identified with (\ref{eig}) with different choices of analytic
continuations over the real axis. This is not an unreasonable idea, as the behavior near $y_1=\pm 1$ of
$\psi_{\omega}$ is closely related to behavior of the solutions of
the field equations near the cosmological horizon.

An important question is what the decomposition of $L^2(S^{d-1})$
in terms of modes living on the two static patches (the northern
and southern diamonds) looks like. Naively, one might be inclined
to simply cut the $d-1$-sphere into two pieces, since this is what
happens at the $t=0$ slice in global coordinates. However, this is
too naive, as the map between functions on $S^{d-1}$ and solutions
of the equations of motion is more complicated than this. The map
sent $\sum c_k Y_k $ to $\sum c_k f_k(t) Y_k$, and even at $t=0$
the values of $f_k$ are highly non-trivial. Therefore, cutting up
the sphere is not the right thing to do. It would be nice to have
a detailed derivation of the right way to decompose things, but
for now we merely present a guess. The guess is that the functions
on $S^{d-1}$ that correspond to modes on the northern diamond are
precisely the eigenfunctions (\ref{eig}) with $\omega>0$, whereas
the ones on the southern diamond are the ones with $\omega<0$.
This would fit in nicely with our proposed picture of de Sitter
complementarity in Sec. 3.3.1.

As a first check of this idea, one would like to see the thermal
nature of the static patch appear. For this, we use a rather
peculiar basis of functions on $S^{d-1}$, namely
\be \label{pecbas} h_k(y_i) = (1-y_1)^{1/2-h^-} (iy + \sqrt{1-y^2})^k
\chi(\omega_{d-2})
\ee
for some integer $k$. If we integrate this basis function $h_k$
against $\psi_{\omega}(y_i)$, we run into the integral
\be \label{inttt}
S_k(\omega) = \int_{-1}^1 dy_1 \frac{1}{\sqrt{1-y_1^2}} \left(
\frac{1+y_1}{1-y_1} \right)^{i\omega/2} (iy + \sqrt{1-y^2})^k .
\ee
One can show, using contour deformation, that for $k\neq 0$
\be \label{ee45}
S_k(-\omega) = (-1)^{k-1} e^{\omega \pi} S_k(\omega) .
\ee
The reason we had to introduce the peculiar basis (\ref{pecbas})
is that in this way we isolate the thermal factor $e^{\omega\pi}$
up to the phase $(-1)^{k-1}$. Up to some subtleties involving this
minus sign, we indeed see that decomposition gives the required
thermal behavior. This is evidence that the decomposition in the
two static patches is simply the decomposition in states with
$\omega>0$ and $\omega<0$.

We leave a more detailed study of this proposal and its
application to the toy model given in Sec. 3 to future work.

\subsection{Example: massless field in two dimensions}

Finally, we illustrate some of the above observations 
by looking at a massless field in two dimensional de Sitter space.
We begin by summarizing the coordinate systems on $-X_0^2+X_1^2 +
X_2^2=1$.

\paragraph{Global coordinates: }
These obey
\be
X_0  =  \tan \tau ~~~;~~~ 
X_1  =  \cos\phi /\cos \tau ~~~;~~~
X_2  =  \sin \phi/\cos \tau,
\ee
with metric
\be
ds^2 =\frac{1}{\cos^2 \tau} (-d\tau^2 + d\phi^2).
\ee
The $SL(2,R)$ action is
\bea
L_0 & = & -2 \cos \tau \cos \phi \partial_{\tau} + 2 \sin\phi \sin\tau
\partial_{\phi} \nonumber \\
L_{1}+L_{-1} & = &
2 \cos \tau \sin \phi \partial_{\tau} + 2 \cos\phi \sin\tau
\partial_{\phi} \nonumber \\
L_{1}-L_{-1} & = & 2 \partial_{\phi} .
\eea
The Euclidean modes are the modes $\exp(im \phi - i |m|\tau)$,
with $m$ an integer. One immediately verifies that $SL(2,R)$ maps
these modes into themselves, and that this gives a representation
of the form (\ref{gencc2}) on $L^2(S^1)$.

\paragraph{Static coordinates: }
These obey
\be
X_0  =  -\sinh t/\cosh\rho  ~~~;~~~
X_1  =  \cosh t /\cosh \rho ~~~;~~~
X_2  =  \tanh\rho,
\ee
with metric
\be ds^2 =\frac{1}{\cosh^2 \rho} (-dt^2 + d\rho^2).
\ee
The $SL(2,R)$ action is
\bea
L_0 & = & 2 \partial_t \nonumber \\
L_{1}+L_{-1} & = &
- 2 \cosh t \sinh \rho \partial_{t} - 2 \cosh\rho \sinh t
\partial_{\phi} \nonumber \\
L_{1}-L_{-1} & = & 2 \sinh \rho \cosh t \partial_{t} +
2 \cosh{\rho} \cosh t \partial_{\phi}
\eea
The other static patch is obtained from this one by sending
$t\rightarrow t+\pi i$.

\paragraph{Inflating coordinates: }
These obey
\bea
X_0 & = & \frac{1}{2}(-\eta^{-1}+\eta -x^2 \eta^{-1})
\nonumber \\
X_1 & = & \frac{1}{2}(\eta^{-1}+\eta -x^2 \eta^{-1})
\nonumber \\
X_2 & = & x\eta^{-1}
\eea
with metric
\be ds^2 =\frac{1}{\eta^2} (-d\eta^2 + dx^2).
\ee
The $SL(2,R)$ action is induced from the other two via the
coordinate transformations below. We only notice that
\be L_0 = -2 \eta \partial_{\eta} -2 x \partial_x .
\ee

The three coordinate systems are related via the following light-cone
maps
\be (\eta \pm x)^2 = e^{-2(t \pm \rho)} = \frac{ 1+\sin(\tau
\mp \phi)}{1-\sin (\tau \mp\phi)} \ee
which is what we would expect for a massless field in
two dimensions.  Since the static coordinates on the southern and northern diamonds
are related via $t\rightarrow t+\pi i$, it is clear that a global
mode which is continued from north to south with behavior
$\exp(i\omega t)$ picks up a factor of $\exp(\pm \pi \omega)$,
which shows the thermal nature of the vacuum defined with respect
to these modes.

In global coordinates, the modes induced from the static patch are
\be \label{eig2}
\Phi^{\pm}_{\omega} = \left( \frac{ 1+\sin(\tau \mp \phi)}{1-\sin
(\tau \mp\phi)} \right)^{2i\omega} .
\ee
These modes obey
\bea
L_0 \Phi^{\pm}_{\omega} & = & -8 i \omega \Phi^{\pm}_{\omega}
\nonumber \\
L_1 \Phi^{\pm}_{\omega} & = & \mp 4 i \omega
\Phi^{\pm}_{\omega-i/4} \nonumber \\
L_{-1} \Phi^{\pm}_{\omega} & = & \pm 4 i \omega
\Phi^{\pm}_{\omega+i/4}  .
\eea
The modes (\ref{eig2}) are indeed of the form given in
(\ref{eig}), once we take $\tau=-\pi/2$, rescale $\omega$, and
identify $y_1$ with $\cos\phi$.

In the global patch the quantization is exactly as for the closed
string. The conformal factor drops out of the action and also out
of the KG inner product. Let us look for a state that is invariant
under $SL(2,R)$.    If $L^l_k$ are the standard Virasoro generators
for the left movers, and $L^r_k$ the standard Virasoro generators
for the right movers, then the $SL(2,R)$ we want to preserve is
\bea
L_1=L^l_1 + L^r_{-1},\qquad L_{-1}=L^l_{-1} + L^r_1,\qquad
L_0=L^l_0-L^r_0
\eea
This is structure similar to what we see for a boundary state, and
we can easily write down the relevant $SL(2,R)$ invariant state in
some form $\exp(\sum_n a^l_{-n} a^r_{-n}) |0\rangle$. This
illustrates the analogy between boundary states, and the $SO(d,1)$
invariant vacuum states that we find in de Sitter space. Perhaps
the one-parameter family of de Sitter invariant vacua is closely
related to the one-parameter family of boundary states for a free
scalar field predicted a long time ago by Friedan, see also recent
paper \cite{boundary}.

Finally, we notice that if we compute the Klein-Gordon inner
product between Euclidean modes and the $L_0$ eigenmodes in
(\ref{eig2}), we run into integrals precisely of the form
(\ref{inttt}). This is further evidence for the idea that the
static patches are simply distinguished by the sign of $\omega$,
as we suggested above.

\section{Conclusions}

In this paper we have explored various physical and mathematical problems inspired by possible holographic descriptions of de Sitter space.      First, in view of the similarities between de Sitter space and Euclidean anti-de Sitter space, we studied to what extent data on one can be mapped onto the other.   We showed that there is a nonlocal map that commutes with the de Sitter isometries and transforms the bulk-boundary propagator and solutions of the free wave equation in de Sitter onto the same quantities in Euclidean anti-de Sitter.   This map also transforms the two de Sitter boundaries into the single Euclidean AdS boundary via an antipodal identification as advocated in \cite{andyds}.   This raises the possibility that the holographic dual to EAdS could also describe de Sitter, but  since the map has a nontrival kernel, our study suggests instead that a holographic dual to de Sitter would involve independent (but possibly entangled) CFTs associated with both de Sitter boundaries.   We reached a similar conclusion by studying the action as a functional of boundary data  for classical scalar fields in dS, as well as for 3d de Sitter gravity in both the Einstein and Chern-Simons formulations.    As part of the exploration we displayed a family of solutions to 3d gravity with a positive cosmological constant in which the equal time sections are arbitrary genus Riemann surfaces.

In the second part of the paper we argued that if de Sitter space is dual to a Euclidean CFT, then the field theory would have novel hermiticity conditions, and realize a different form of conformal symmetry that we called the {\it Euclidean Virasoro algebra}.   Since there is no concrete proposal for a field theory dual to de Sitter, we explored the Euclidean Virasoro symmetry in the context of a free boson and went a long way towards establishing properties of such theories.   In addition to exploring unitary realizations of the symmetry, the structure of the spectrum and partition function, we discussed how we anticipate the novel CFTs to play a role in describing de Sitter space.  For example, we showed that our new hermiticity conditions it is possible to construct an $SL(2,C)$ invariant state in a product of two $SL(2,C)$ invariant theories.  Such a state is a candidate for a dual description of the de Sitter vacuum.  We also argued it might be possible to rewrite the product of two such Hilbert spaces realizing Euclidean Virasoro symmetries, as a product of highest and lowest weight representations.   Neither of the latter would realize the $SL(2,C)$ symmetry of de Sitter space unitarily but this split would be very natural for describing the physics in antipodal static patches of de Sitter space, with thermal entanglement between the highest/lowest weight factors giving rise to de Sitter thermodynamics.    Finally we provided evidence for this picture by examining the physics of a scalar field in de Sitter space, explaining how the dS isometries are realized on it and how mode solutions in the global and static patches are related.

Our picture is also able to accomodate other asymptotically de Sitter spaces.  For example, the spinning conical defects in 3d de Sitter \cite{deserjackiw,park} would simply correspond to different entangled states.  We would expect these states to have lower entropy than de Sitter when decomposed into representations appropriate to antipodal static patches.   It is possible that thinking about the representation theory of such entangled states that preserve the asymptotic de Sitter symmetries would shed light on the de Sitter mass bound conjecture of \cite{bdbmds}.    If there is a dS/CFT correspondence, there is a relation between time evolution in spaces with a positive cosmological constant and RG flow in the dual field theory \cite{rgstrom,bdbmds,rgothers}.   In our picture, any asymptotically dS space would correspond to an entangled stated in a suitable product of CFTs, and we would compute a flow in this state to holographically describe time evolution.    In general, if we compute the quantum gravity path integral between specified asymptotically de Sitter boundary conditions we would expect to sum over all intermediate geometries, including singluar and topologically disconnected ones.   Possibly these additional spaces have a relation to the mass bound conjecture of \cite{bdbmds} and appear in the dual CFT spectrum above the de Sitter bound.

It has been suggested that there cannot be a dS/CFT correspondence at all \cite{susskind}. The argument in \cite{susskind} started with dS space with a cutoff in time, and with two operators, one on each  boundary, and each part of the same static patch. As we take the boundaries to infinity, quantum gravity
corrections will induce a finite tail in the two-point function, and therefore the asymptotic scaling behavior needed for a dS/CFT correspondence does not exist, according to \cite{susskind}.

This quantum gravity calculation is obviously hard to do. In \cite{susskind}, the finite tail is obtained by assuming that we are computing in a thermal ensemble with a discrete spectrum. Since the relevant quantum gravity calculations probe the full global structure of de Sitter space, we  need to assume that global de Sitter is described by a thermal state in a theory with a discrete spectrum. In our setup, however, global de Sitter is described by a pure state, and the finite entropy is due to entanglement.  Therefore, if we do the calculation suggested in \cite{susskind}, we will most likely not find a finite tail. There are complicated, non-local operators in the dual theory that do probe the finite tail and the thermal nature of the static patch. These are obtained by tracing over a suitable part of the Hilbert space, for which a preliminary proposal was given in section~4.3. It is these operators that describe the static patch, but they are not the conventional local operators in the boundary theory.

This work has an exploratory character and there are many loose ends and potential directions forward.  For example, if CFTs exist with the novel hermiticity conditions that we have described, can they be used to define new string theories?  We describe this possibility in greater detail below and then proceed to enumerate a number of worthwhile directions for future work.

\subsection{New string theories?}

Can there exist new string theories based on non-canonical reality conditions discussed above, ignoring for the moment the problems with the spectrum and partition function in the toy model discussed in Sec.~3?   One possibility is that such string theories underlie cosmological spacetimes like de Sitter.  In this regard it is interesting to investigate the  hermiticity conditions in a theory where we T-dualize a time direction as proposed by Hull \cite{hull}. We leave the detailed study of this question for future work, but here we give an heuristic argument that  new reality conditions can give imaginary B-fields as in Hull's theories.

First, we change the hermiticity from $L_n^{\dagger} = -\bar{L}_n$ to $L_n^{\dagger}=\bar{L}_{-n}$. This
gives a theory with $SL(2,C)$ symmety rather than $SL(2,R) \times SL(2,R)$. Notice that this is the usual hermiticity condition of string theory, except that right and left movers are interchanged. Normally, the gravition and B-field vertex operators are
\be
\epsilon_{\mu\nu} \alpha_{-1}^{\mu} \bar{\alpha}_{-1}^{\nu}
| k \rangle
\ee
and this is a real field, because the complex conjugate (not the
hermitian conjugate) of $\alpha_{-1}$ is itself. The new hermiticity
conditions are the same as the old ones, except that left and
right movers are also interchanged in the process. It is reasonable
that the same is true for complex conjugation. Hence, the vertex
operator above goes under complex conjugation over into
\be
\epsilon_{\nu\mu}^{\ast}
\alpha_{-1}^{\mu} \bar{\alpha}_{-1}^{\nu}
| k \rangle .
\ee
In the usual case we had $\epsilon_{\mu\nu}^{\ast}$ as prefactor,
and both the graviton and antisymmetric tensor have a real
polarization. In the present case, the graviton polarization still
is real, but the B-field polarization is imaginary, exactly as in Hull's theories which contain de Sitter space.

Even though this observation is hardly a proof that we need non-canonical
hermiticity conditions in order to construct string theories of asymptotically
de Sitter spaces (and perhaps more general time-dependent backgrounds), this idea warrants further investigation.   Other recent interesting attempts to find de Sitter space in string theory involve non-critical string theories \cite{eva,alex} and spacelike branes \cite{spacelike}.

\subsection{Open Problems}

Given the exploratory nature of this work it is appropriate to conclude with a list of open problems raised by our studies:

\begin{itemize}

\item Study the extension of the nonlocal map from dS to EAdS to the full interacting supergravity.

\item Explore other realizations of the non-canonical hermiticity conditions to find examples in which the spectrum and partition function are fully under control.

\item Explicitly show how unitary representations of the Euclidean Virasoro algebra based on the principal series for $SL(2,C)$ can be decomposed into products of highest and lowest weight Virasoro representations, as expected from de Sitter physics.  Likewise study whether  entangled states in a product of highest and lowest weight representations can lead to a 1-parameter family of $SL(2,C)$-invariant states in a product of two unitary Euclidean Virasoro representations, or whether there is a unique $SL(2,C)$-invariant entangled state corresponding to the Euclidean vacuum of de Sitter space.

\item Explain de Sitter thermodynamics using the novel dual CFTs introduced here.

\item
Explore the extension of the structures discussed in this paper to $D$ dimensions and to other time dependent backgrounds.  In particular study whether and how time can be holographically generated in cosmological backgrounds via RG flows of a dual Euclidean theory \cite{bdbmds,rgstrom,rgothers}.

\item Study string theories based on CFTs with the new hermiticity conditions.   Do they naturally arise from T-duality in time as in Hull's work \cite{hull}?    Can such string theories apply to more general time-dependent backgrounds?

\end{itemize}

Of course, the single  most important open problem in this area is to find de Sitter space or any other expanding universe  as a controlled solution to string theory.   A concrete stringy realization of such spaces and their holographic duals would teach us a great deal about the role of time in quantum gravity.

\vspace{.5in}

{\bf Acknowledgments: }
We have benefitted from conversations with many colleagues including Tom Banks, Per Berglund, Robbert Dijkgraaf, Willy Fischler, Jim Gates, Tristan Hubsch, Antal Jevicki, Shamit Kachru, Esko Keski-Vakkuri, Per Kraus, David Kutasov, Finn Larsen,  Rob Leigh, Juan Maldacena, Rob Myers, Simon Ross, Eva Silverstein, Andy Strominger, Erik Verlinde, Herman Verlinde, and Edward Witten.    {\small VB} is supported by  DOE grant DE-FG02-95ER40893.    {\small JdB} was also supported in part by NSF grant PHY-9907949 at ITP, Santa Barbara.   We enjoyed the hospitality of many institutions while this work was being completed, including the University of Pennsylvania (JdB and DM), KIAS (VB and DM), the CIT-USC Center for Theoretical Physics (DM) and Humboldt University, Berlin (DM).


\end{document}